\documentclass[fleqn,usenatbib]{mnras}

\usepackage{newtxtext,newtxmath}

\usepackage[T1]{fontenc}

\DeclareRobustCommand{\VAN}[3]{#2}
\let\VANthebibliography\thebibliography
\def\thebibliography{\DeclareRobustCommand{\VAN}[3]{##3}\VANthebibliography}

\usepackage{graphicx}	
\usepackage{amsmath}	

\usepackage{gensymb}
\usepackage[dvipsnames]{xcolor}
\usepackage{fix-cm}
\definecolor{changes}{RGB}{0,0,0} 

\definecolor{notes}{RGB}{49, 135, 12} 

\usepackage{enumitem}

\setlist[enumerate]{leftmargin=*, label=(\roman*)}
\setlist[itemize]{leftmargin=*}

\title[Internal + external effects on planet systems]{The impact of internal versus external perturbations on close-in exoplanet architectures}
\author[Schoettler, C. and Owen, J. E.]{Christina Schoettler$^{1,2}$\thanks{E-mail: c.schoettler@keele.ac.uk} and James E. Owen$^{2}$
\\
$^{1}$Astrophysics Group, Keele University, Keele ST5 5BG, UK \\
$^{2}$Imperial Astrophysics, Department of Physics, Imperial College London, Prince Consort Rd, London SW7 2AZ, UK}

\date{Accepted XXX. Received YYY; in original form ZZZ}

\pubyear{\the\year{}}

\begin{document}
\label{firstpage}
\pagerange{\pageref{firstpage}--\pageref{lastpage}}
\maketitle

\begin{abstract}
Young planetary systems are subjected to different dynamical effects that can influence their orbital architecture over time. In systems with more than one planet, other planets can internally influence each other, for example, through planet-planet scattering. External perturbing effects also need to be taken into account, as stars do not form by themselves but together with other stars in high-density young star-forming regions. This birth environment can externally affect young multi-planet systems through stellar fly-bys. Previous work has shown that the absence/presence and location of an outer giant planet around a close-in planet system do not change how these inner planets react to a single fly-by with another star. We further explore this by comparing the effects of these external perturbations on four close-in sub-Neptune planets to those caused by a situation where only the distant giant is perturbed by the same kind of encounter. Our results indicate that the close-in planet systems have a ``preferred'' end state after 500 Myr, which is reached regardless of how it was perturbed. In addition, the mass of the giant appears not to impact the reaction of the inner planet system in the scenario of an external perturbation in our tested set-ups, i.e. either a single 1 or 5 M$_{\rm{Jup}}$ giant placed at 2.5, 5, 10 or 20 au. However, the mass affects the subsequent evolution of the inner planets if only internal perturbations by the giant are considered. The reduction in mass leads to an absence of collisions during the 500 Myr.

\end{abstract}




\section{Introduction} \label{Intro}

Stars and planets have been shown to form at virtually the same time \citep[e.g.][]{2020ApJ...904L...6A}. Thus, any environment to which the stars are exposed during this time will inevitably also affect the planets during their formation process in the protoplanetary disc \citep[e.g.][]{2015ApJ...809...93D,2020Natur.586..228S,2020ApJ...891..166W,2020RSOS....701271P,2023ASPC..534..685P,Qiao2023}. One of the key aspects of the environmental impact is that stars are born with other stars in grouped environments, e.g. young star-forming regions. These environments are often of higher density than what a star-planet system will experience once the birth region has dispersed into the field \citep[e.g.][]{RN25,RN59}.  Along with these, comes an increased chance for stellar encounters \textcolor{changes}{within a star-forming region}, which can influence the orbits of any exoplanet system around its host stars \citep[e.g.][]{1984AJ.....89.1559H,2006ApJ...641..504A,2009ApJ...697..458S,2011MNRAS.411..859M,2017MNRAS.470.4337C,2024ApJ...970...97Y,2025A&A...696A.175C}. \textcolor{changes}{These regions also show a lower velocity dispersion than the field, resulting in longer encounters, which makes the effect they have more pronounced \citep{2015MNRAS.449.3543W}}.

In the past two decades, many advances have been made in discovering planets outside our solar system. Although the architectures of these \textcolor{changes}{planetary systems} have shown a multitude of different structures, there are a few common aspects that repeatedly appear in observed exoplanet systems. Close-in (periods $\lesssim 100$~days) super-Earth/sub-Neptunes are common \citep[e.g.][]{2013ApJ...766...81F,2015ApJ...799..180S}. Furthermore, they appear to show a correlation with the presence of a more distant outer giant planet \citep[e.g.][]{2016ApJ...821...89B,2019AJ....157...52B,2018AJ....156...92Z,2021A&A...656A..71S,2024MNRAS.528.7202G,2025arXiv250106342V}. These distant giants are often undiscovered in transit observations, but they can dynamically influence their inner planetary systems \citep[e.g.][]{2017AJ....153...42L,2017MNRAS.467.1531H,2017MNRAS.469..171R,2017MNRAS.468.3000M,2018MNRAS.478..197P,2019MNRAS.482.4146D,2020MNRAS.498.5166P}. This can happen when the giant planet is perturbed on to an inclined or eccentric orbit or if it becomes otherwise dynamically unstable. These perturbations can be the result of dynamical encounters with other stars \citep[e.g.][]{2012MNRAS.419.2448P,2024ApJ...970...97Y}, but can also be due to internal planet-planet scattering, \citep[e.g.][]{2008ApJ...686..621F, 2017MNRAS.470.4337C, 2020MNRAS.491.1369A}, or interactions with the protoplanetary disc \citep[e.g.][]{2017A&A...598A..70S, 2018MNRAS.474.4460R, 2023MNRAS.523.2832R}.

Young star-forming regions have a wide variety of different densities, creating different encounter rates. The presence of other stars has the potential to provide an external source of perturbation for infant planet systems, affecting both the distant giant and the smaller inner planets \citep[e.g.][]{2022MNRAS.509.1010R}. \textcolor{changes}{ By comparing to N-body simulations of star forming clusters with realistic distributions \citet{2024MNRAS.533.3484S}. found that 10-20 per cent of sun-like stars likely experienced a close encounter at $<300~$pc. }

\begin{figure*}
    \centering
    \includegraphics[width=1.0\linewidth,trim={2.1cm 16cm 1.1cm 2.7cm},clip=true]{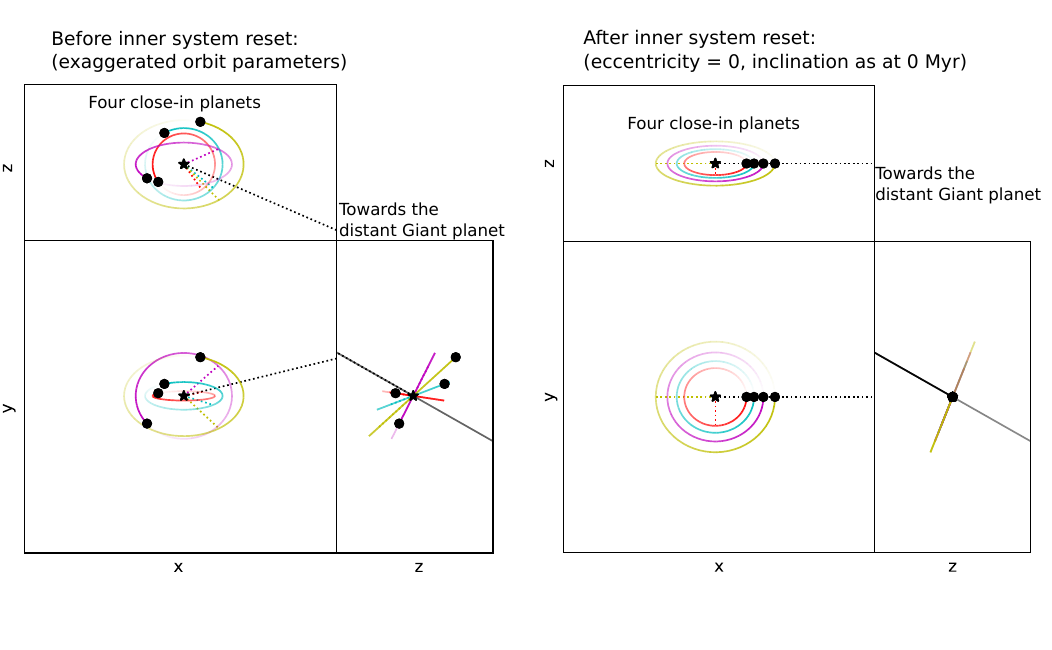}
    \vspace{-15mm}
    \caption{Schematic showing the reset approach applied for the close-in four-planet system. The left graphic shows ``exaggerated'' orbits of the four planets after the encounter when they are on slightly eccentric, non-coplanar orbits with the giant planet outside the axis limits. Note: In our simulations, none of the planets show this amount of perturbation at the reset at 1 Myr. On the right graphic after the reset, the four-planet system is set back to its original circular, coplanar orbits, but the giant planet's orbit remains unchanged, being eccentric and/or inclined as shown in Fig. \ref{fig:Inc_ecc_GP_only_10_14_lowmass}. The inner planet system is reset to its initial inclination at the start of the simulations to isolate the effect of internal perturbations after 500 Myr.}
    \label{fig:schematic_reset}
\end{figure*}

Internal effects by large giant planets on inner sub-Neptunes or super-Earths can range from excitation of their mutual inclinations \citep[e.g.][]{2017AJ....153...42L,2017MNRAS.467.1531H,2021MNRAS.502.3746R} and/or their eccentricities \citep[e.g.][]{2017MNRAS.468.3000M,2018MNRAS.478..197P,2019MNRAS.482.4146D,2020MNRAS.498.5166P} to affecting the formation processes of the inner planets itself \citep[e.g.][]{2024A&A...687A.121K}.

\citet{2022MNRAS.509.1010R} evaluated the impact of one or more externally perturbed outer planetary companions on the orbits of inner planet systems. They suggested that the perturbations of the giant planets caused by the encounter could then trickle to the inner planets and destabilise their orbits. They find that more than one outer giant is required to perturb the inner system considerably. However, they did not allow a long period of time for these perturbations to filter towards their inner planet system. They ran their simulations only for a short time, which at most is 100 times the orbital period of their giant planet, i.e. a few Myr. \textcolor{changes}{\citet{2020MNRAS.496.1149L} also investigated encounters on close-in planet and giant planet systems and found that no planets were lost during their 1 Myr simulation run (for their Kepler-48 system) after a fly-by with a closest encounter distance beyond 4 au. In both instances, running the simulations for several hundred Myr could have shown that a destabilisation could also occur for a single outer giant or close fly-by; \citet[][]{2011MNRAS.411..859M} have shown that it can take a long time for perturbations to have an effect.} 

There has been limited work comparing internal and external perturbation effects on close-in-planet systems; however, some have looked at these effects on Trans-Neptunian Objects \citep[e.g.][]{2020CeMDA.132...12S}. They find that there is a distance component to the perturbation, which influences how these objects react to the perturbation and whether internal or external perturbations dominate the resulting effect. \citet{2024MNRAS.533.3484S} investigated the dynamical effect that a single stellar fly-by, which can commonly occur in the star's birth cluster, will have on a close-in exoplanet system, either with or without a distant giant planet. They found that the level of perturbations at 500 Myr, as measured either by mutual inclination differences or the number of remaining inner planets, was not significantly different for any of the tested placements of the giant planet's initial semi-major axis (5, 10 or 20 au) or lack of a giant planet. This suggested that the fly-by alone could be responsible for a perturbed close-in planet system and would not require a perturbed outer giant at all; that is, the ``external'' perturbation from the stellar flyby dominates the long-term outcome.

We now take this work further and investigate this result in more detail. Compared to \citet{2024MNRAS.533.3484S}, we add a further distance and also include lower-mass giant planets. In addition, we completely remove the effect of the fly-by from the close-in planet system to allow us to compare the internal perturbations caused by the giant that was affected by the fly-by with systems where all planets have been affected by a single fly-by. Although this is not a physical scenario, it is useful to help us diagnose the relative importance of internal and external perturbations.  

In section \ref{method}, we describe the reset procedure we implemented to remove the external perturbation from the inner planets and additional aspects of the simulation set-up. 
In section \ref{results_giants}, we briefly evaluate the perturbation of the giant planet as a result of the fly-by, whereas in section \ref{results}, we focus on the results of our simulations comparing internal and external perturbations and discuss them in section \ref{discussion}. Finally, we summarise our work in section \ref{conclusion}.
 
\section{Method}\label{method}

One of the advantages of \citet{2024MNRAS.533.3484S} was to base the fly-bys on the properties of those that can occur in typical star-forming regions, such as the Orion Nebula Cluster. These fly-bys were shown to occur closer than 300 au for more than 50 per cent of all encounters in their simulations. In this present analysis, we compare the effect of the same kind of fly-by encounters with other stars on the orbits of close-in multi-planet systems in the presence/absence of a distant giant planet. We separate the effect into ``internal'' perturbations: those arising from the perturbed giant planet on the close-in planets; and ``external'' perturbations: those from the fly-by directly on the close-in planets (and their distant giant companion). 

\begin{figure*}
    \centering
    \begin{minipage}[t]{0.9\columnwidth}
         \centering
    	\includegraphics[width=1.0\linewidth]{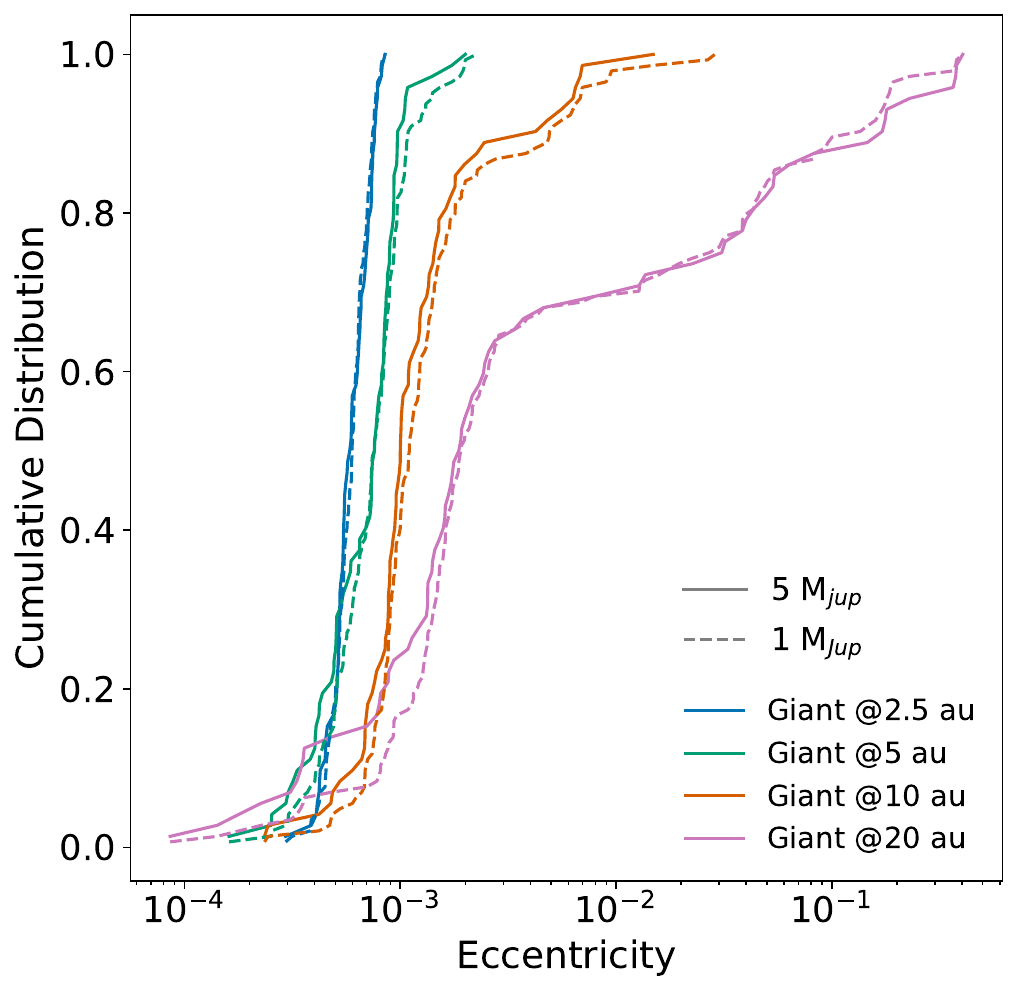}
     \end{minipage}
        \centering
        \vspace{0pt}
    \begin{minipage}[t]{0.9\columnwidth}
    	\includegraphics[width=1.0\linewidth]{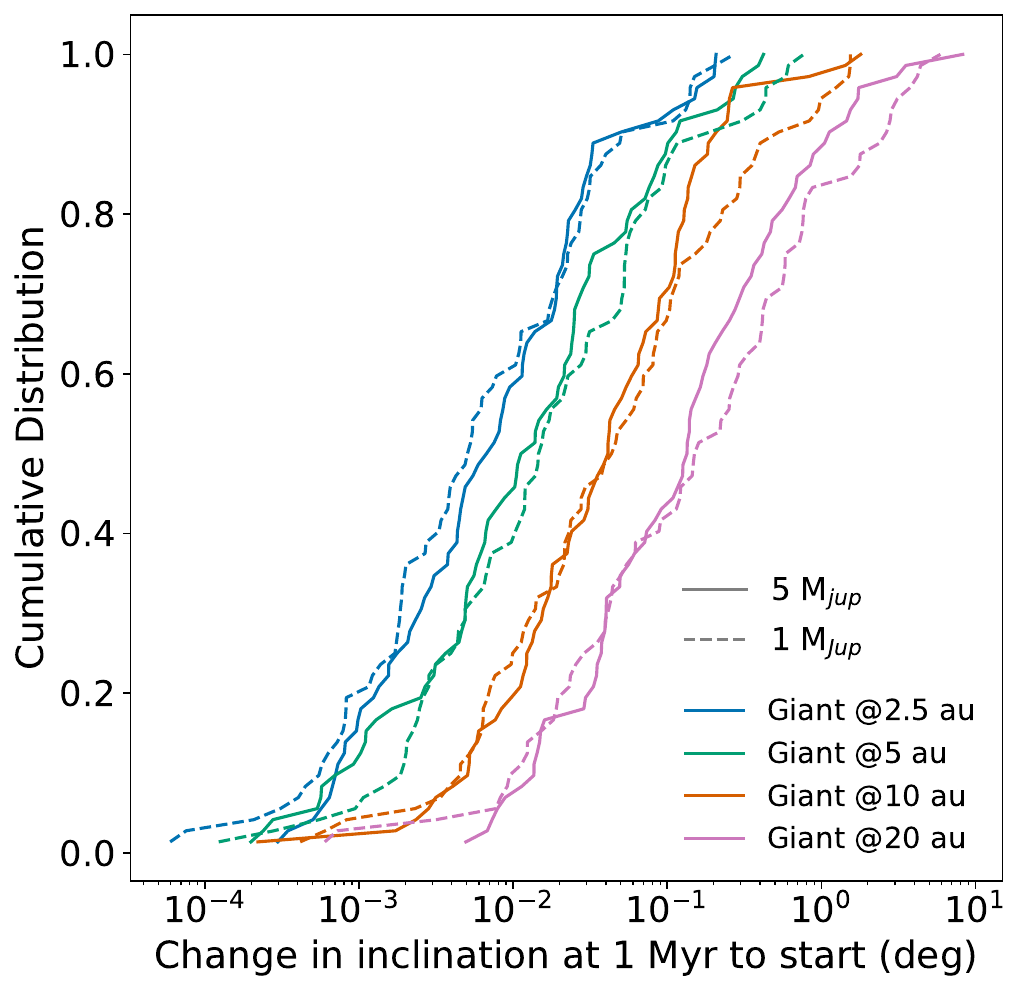}
      \end{minipage}
     \caption{Comparison of the cumulative distributions at 1 Myr of the changes in orbital parameters of the higher-mass giant planet ('solid`) and lower-mass giant planet ('dashed`) in reaction to the fly-by (fly-bys occur within the first 20kyr). \textbf{Left:} Eccentricity for the two giant planet masses of all 10+14 R$_{\rm{MH}}$ (mutual Hill radii) simulations combined at 1 Myr. All giant locations show very similar cumulative distributions of their eccentricity.  \textbf{Right:} Change of the inclinations for the two giant planet masses in all simulations at 1 Myr compared to the start. The cumulative distributions show similar shapes for all four giant planet locations, indicating a similar level of increase in inclination regardless of mass.}
     \label{fig:Inc_ecc_GP_only_10_14_lowmass}
\end{figure*}

Our ``fiducial'' planetary systems are initially made up of four close-in sub-Neptunes of 5 M$_{\oplus}$ and a more distant giant planet with a mass of 5 M$_{\rm{Jup}}$ placed at different distances, \textcolor{changes}{as the inner planets are added as sub-Neptunes, we assume an initial radius of 3 R$_{\oplus}$. If they collide with another of the close-in planets, their radius changes and will be determined based on that post-collision mass using the mass-radius relationship for rock/iron planets from Eq. 8 in \citet{2007ApJ...659.1661F,2007ApJ...668.1267F}}. The four close-in planets are placed at two different separations, 10 and 14 \textcolor{changes}{mutual Hill radii (R$_{\rm{MH}}$)}, both stable in isolation without a fly-by (confirmed by direct integration to 500 Myr). \textcolor{changes}{The innermost planet is always located at 0.1 AU, and the outermost (close-in) planet location is determined by the separation choice and are at 0.1915 au and 0.2509 au, respectively}. We use the simulations from \citet{2024MNRAS.533.3484S}, where the giant is placed at 5, 10 and 20 au in initially circular and coplanar orbits, and refer to that source for further details on the setup. We add a fourth location for the giant planet, which will be closer to the inner planets, placed at 2.5 au. At this location, the giant will be further away from our closest fly-by encounter, which occurs at 50 au, but closer to the inner planets. For the fly-by simulations, all planets will be subjected to the encounter after 10--20 kyr, and their subsequent dynamical evolution will then be simulated to an age of 500 Myr. Although the lengths of our simulations are significantly shorter than the typical ages of many known Kepler planetary systems, running our simulations to match average ages, i.e., $\sim$5 Gyr, is technically not feasible, as it would greatly extend the required CPU wall time. However, as we shall find, the dynamics of the systems have settled by 500 Myr, and we can begin to draw conclusions.

We want to separate the internal and external perturbations and do that at the time when we remove the fly-by star at 1 Myr. To investigate the internal perturbations, we want to isolate the effect of the inclined and eccentric giant planet on the close-in planetary system without the external effect. To do this, we replace the perturbed inner four-planet system with the original unperturbed system. In this way, the close-in planetary system will start on coplanar and circular orbits, whereas the giant has been perturbed. To facilitate this, we are fully resetting the inner four planets to their locations and orbital parameters at the start of the simulations, separated by 10 or 14 R$_{\rm{MH}}$ as shown in Fig. \ref{fig:schematic_reset}. This is done for all four locations of the giant planet (2.5, 5, 10 and 20 au). Once again, we follow the dynamical evolution of the planets up to an age of 500 Myr, but this time influenced only by the previously perturbed giant. In this way, we expect to be able to separate the total perturbation into that from the fly-by and that only induced by the giant.

In addition, we also vary the mass of the giant planet by running a completely new set of simulations that are otherwise identical to those described above (and in \citealt{2024MNRAS.533.3484S}). The original mass of our giant planets was set at 5 M$_{\rm{Jup}}$, and here we run a new mass of 1 M$_{\rm{Jup}}$. This will enable us to determine the role that the giant's mass plays in the internal and external perturbations. These two masses straddle the typical extra-solar giant planet mass of a few Jupiter masses \citep[e.g.][]{Fulton2021}. As in \citet{2024MNRAS.533.3484S}, the central star is a 1.0 M$_{\sun}$, and the fly-by characteristics are the same as in \citet{2024MNRAS.533.3484S}. We do not evolve the planets' radii unless they are part of a collision. In that case, the same procedure will be applied by turning the initial sub-Neptunes into super-Earths with radii depending on their post-collision mass.

\begin{figure*}
    \centering
    \begin{minipage}[t]{0.9\columnwidth}
         \centering
        \includegraphics[width=0.95\linewidth]{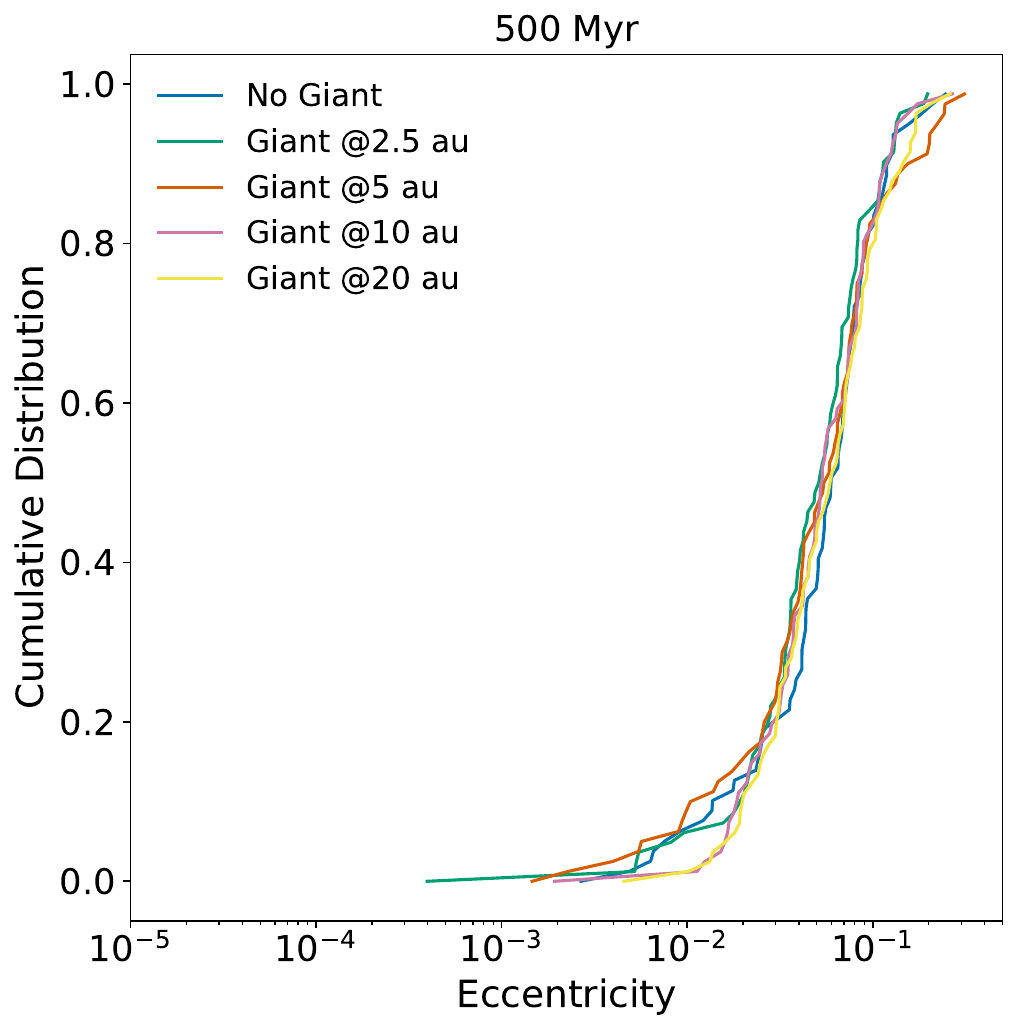}
     \end{minipage}
        \centering
        \vspace{0pt}
    \begin{minipage}[t]{0.9\columnwidth}
    	\includegraphics[width=0.95\linewidth]{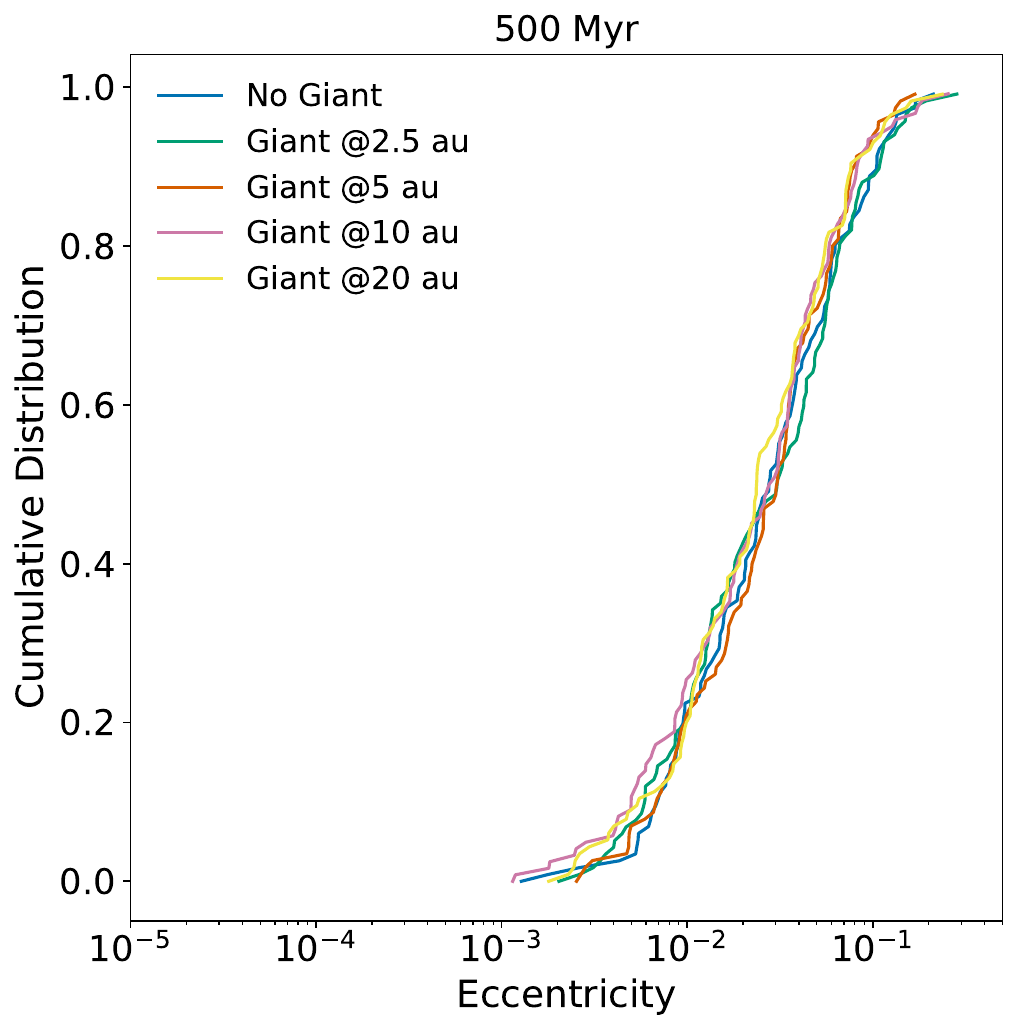}
      \end{minipage}
       \centering
        \vspace{0pt}
     \caption{\textbf{Left:} Comparison of the eccentricities of all remaining close-in planets at 500 Myr for the simulations with an initial separation of 10 R$_{MH}$ for the giant planet with 5  M$_{\rm{Jup}}$. \textbf{Right:} Comparison of the eccentricities of all remaining close-in planets at 500 Myr for the simulations with an initial separation of 14 R$_{MH}$ for the same outer giant mass. Both plots show that for the simulations where all planets were subjected to the fly-bys, the eccentricities of the remaining planets at the end of our simulations are very similar, indicating that the presence/absence and exact location of a distant giant planet make no apparent difference to this orbital characteristic. This is a similar result compared to other orbital characteristics, such as the average semi-major axis and inclination (compared to the start) of the close-in planets (see Appendix Figs. \ref{fig:CDF_all_semax} and \ref{fig:CDF_all_incl})}
     \label{fig:Ecc_evolv_10_14_all_comp}
\end{figure*}

Our previous work showed that the perturbing effect on the inner four planets is independent of their initial orbital starting position. In order to keep computational expense low, we only use one of the two original orbital starting positions, which is that they start lined up for our new simulations\footnote{Although we note that the planet's phases are randomised by the time the fly-by arrives at the system.}. We use the same 36 fly-by parameters, which are made up of three different mass perturbers (0.1, 0.5 and 1.0 M$_{\odot}$ passing with two different velocities (2 and 4 km\,s$^{-1}$) and three different closest encounter distances (50, 150 and 250 au). These 18 perturbers approach host star-planet systems from two different \textcolor{changes}{randomly chosen} directions, resulting in changes in the interaction angles between the perturber and the orbital plane of the planets. The general perturber characteristics, such as encounter distance, are motivated by simulations of the early dynamical evolution of star-forming regions \citep[e.g.][]{2020MNRAS.495.3104S}.

To allow a direct comparison with our previous simulations, we use the REBOUND $N$-body code to integrate these systems. We run 216 fly-by simulations using different fly-by and inner planet separation configurations with the three existing giant planet locations, where the inner planet system is replaced at 1 Myr. We then add 72 simulations of the planet system set-up where the giant is located at 2.5 au, these are run with and without a replaced inner planet system. All of our simulations are run to 500 Myr. The individual timestep snapshots are stored in the REBOUND SimulationArchive with timesteps of 10$^3$ yr during the first 1 Myr using the IAS15 15th order Gauss-Radau integrator \citep[][]{reboundias15,reboundsa}. As in our previous work, we remove the stellar perturber at 1 Myr, change the integrator to the hybrid MERCURIUS integrator \citep{reboundmercurius} and increase the time between stored snapshots to 10$^4$ years. For the simulations where we reset the inner planet system, the first 1 Myr are identical between the simulations for the internal and external perturbation effect impacts. We then use MERCURIUS with the same snapshot timesteps as mentioned for the remaining integrations.

\section{Perturbation of the giant planet by the fly-bys}\label{results_giants}

Before comparing the internal to the external perturbing effects on the four inner planets, we evaluate how the giant planets are affected by the fly-by. We show the evolution of the eccentricity and inclination changes compared to the start of the simulations. The left plot in Fig. \ref{fig:Inc_ecc_GP_only_10_14_lowmass} shows the eccentricity of the different giant planet locations as a cumulative distribution for all simulations after 1 Myr. As previously mentioned, the initial eccentricity of the giant planets is zero, and we find that the planets show non-zero eccentricities at 1 Myr when we replace the close-in planetary system. The simulations with giants initially placed at 2.5, 5 and 10 au show similar overall levels of eccentricity. In contrast, for the 20 au giant placement, the results show that $\sim$30 per cent of the simulations show higher eccentricities than the other three locations. This is true for both tested giant planet masses. The maximum eccentricity is $\sim$0.4 in these 20 au simulations. In the same figure on the right, we see the comparison of the inclination of the giant planets compared to their starting inclination. In an unperturbed system, this can be expected to be zero after our short initial integration time of 1 Myr. However, we find that for all giant planet locations, the inclinations have increased. The closer the giant planet was placed to the perturbing fly-by star, the larger the change. The maximum change in inclination is $\sim$3.1$\degree$ for a 20 au giant planet. Even a few degrees of mutual inclination for the wide-orbit giants is sufficiently large that they would not transit when the inner planets would, and vice versa. 


\begin{figure*}
    \centering
    \begin{minipage}[t]{0.9\columnwidth}
         \centering
    	\includegraphics[width=0.95\linewidth]{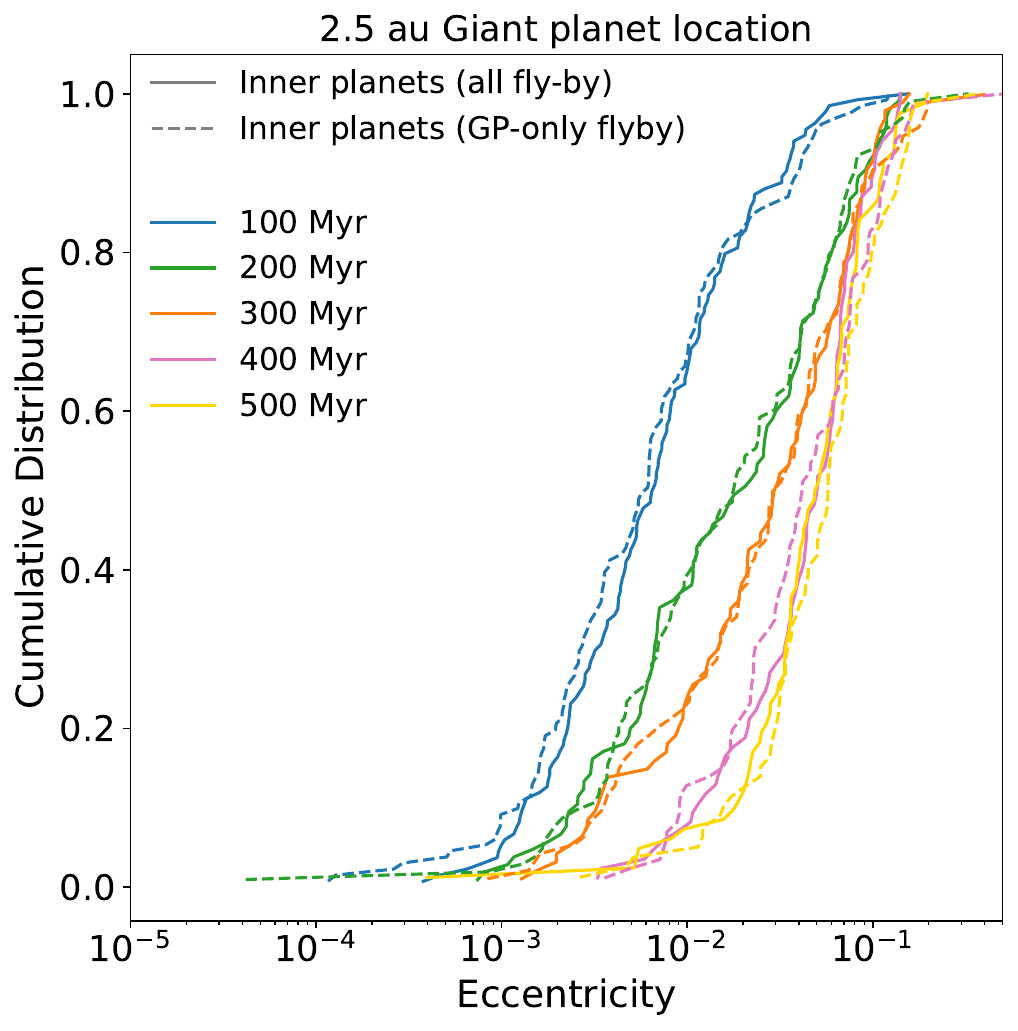}
     \end{minipage}
        \centering
        \vspace{0pt}
    \begin{minipage}[t]{0.9\columnwidth}
    	\includegraphics[width=0.95\linewidth]{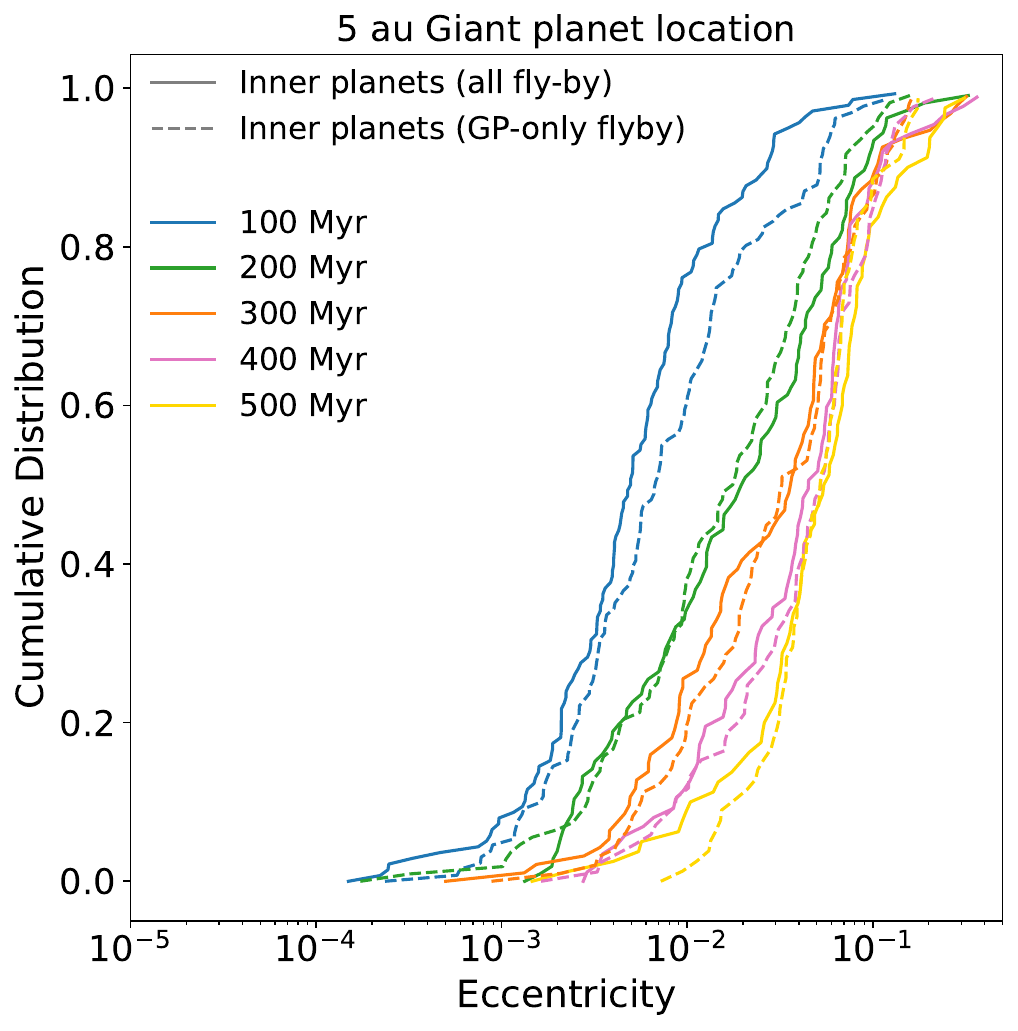}
      \end{minipage}
       \centering
        \vspace{0pt}
    \begin{minipage}[t]{0.9\columnwidth}
         \centering
    	\includegraphics[width=0.95\linewidth]{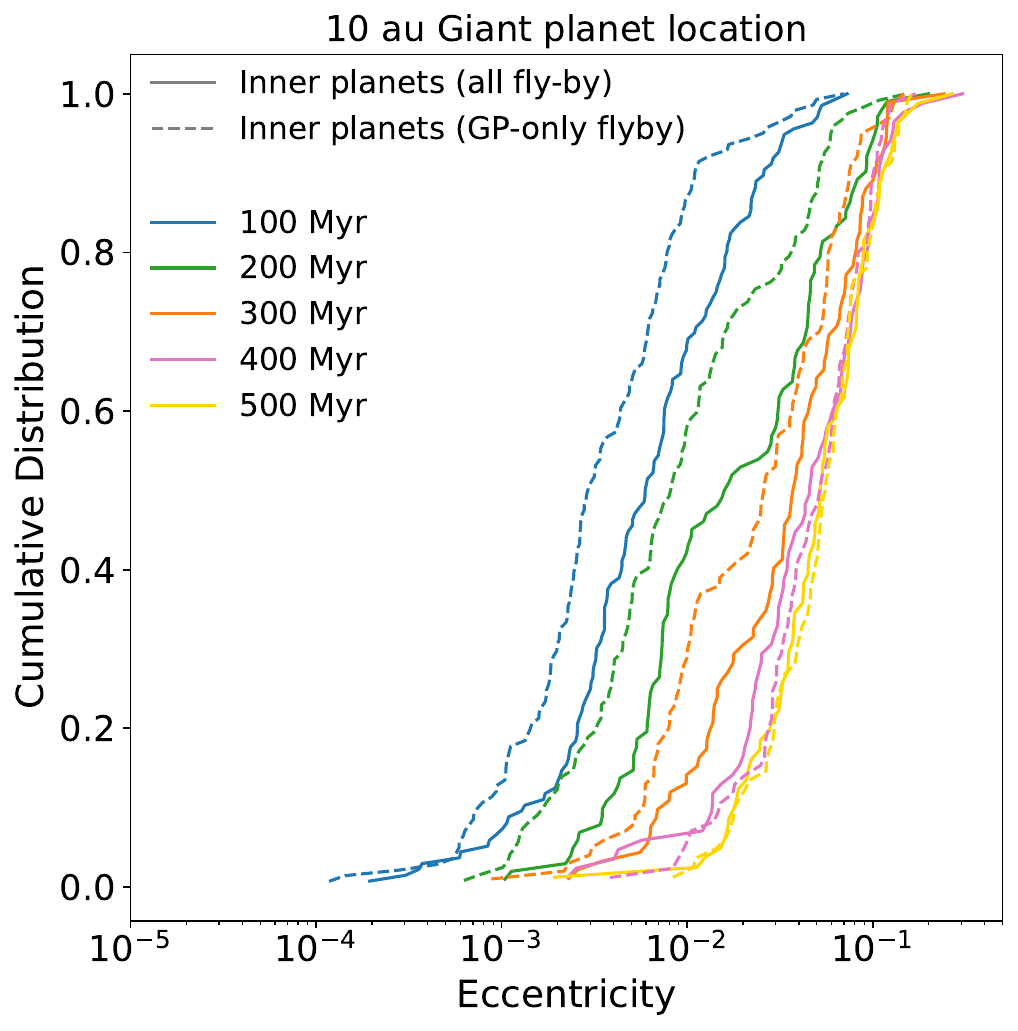}
     \end{minipage}
        \centering
        \vspace{0pt}
    \begin{minipage}[t]{0.9\columnwidth}
    	\includegraphics[width=0.95\linewidth]{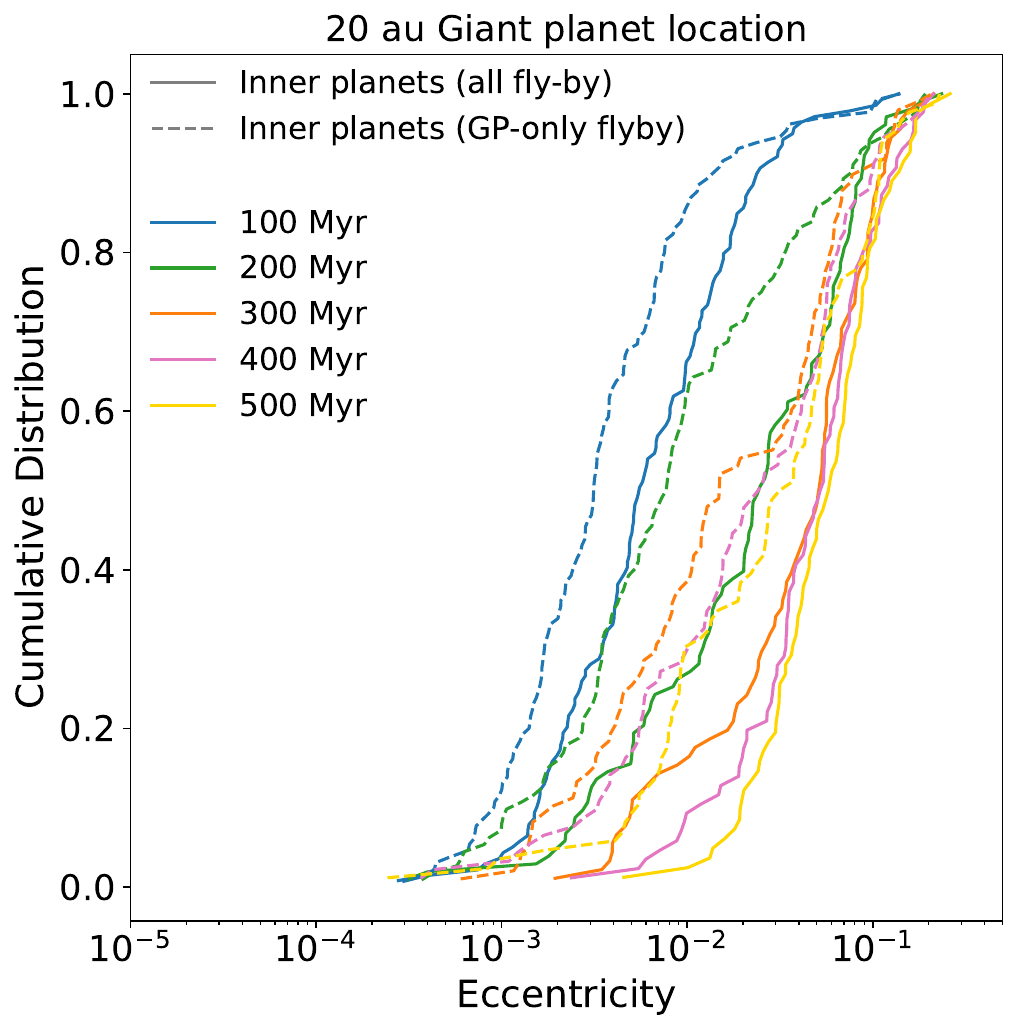}
      \end{minipage}
     \caption{Time evolution of the cumulative distributions of the eccentricity of all remaining inner planets in simulations with an initial separation of 10 R$_{MH}$ for different placements of the giant planet. We compare the eccentricity for the close-in planets subjected to a fly-by ('solid` lines) and those where only the distant giant was subjected to that fly-by ('dashed` lines). }
     \label{fig:Ecc_evolv_10}
\end{figure*}

\begin{figure*}
    \centering
    \begin{minipage}[t]{0.9\columnwidth}
         \centering
    	\includegraphics[width=0.95\linewidth]{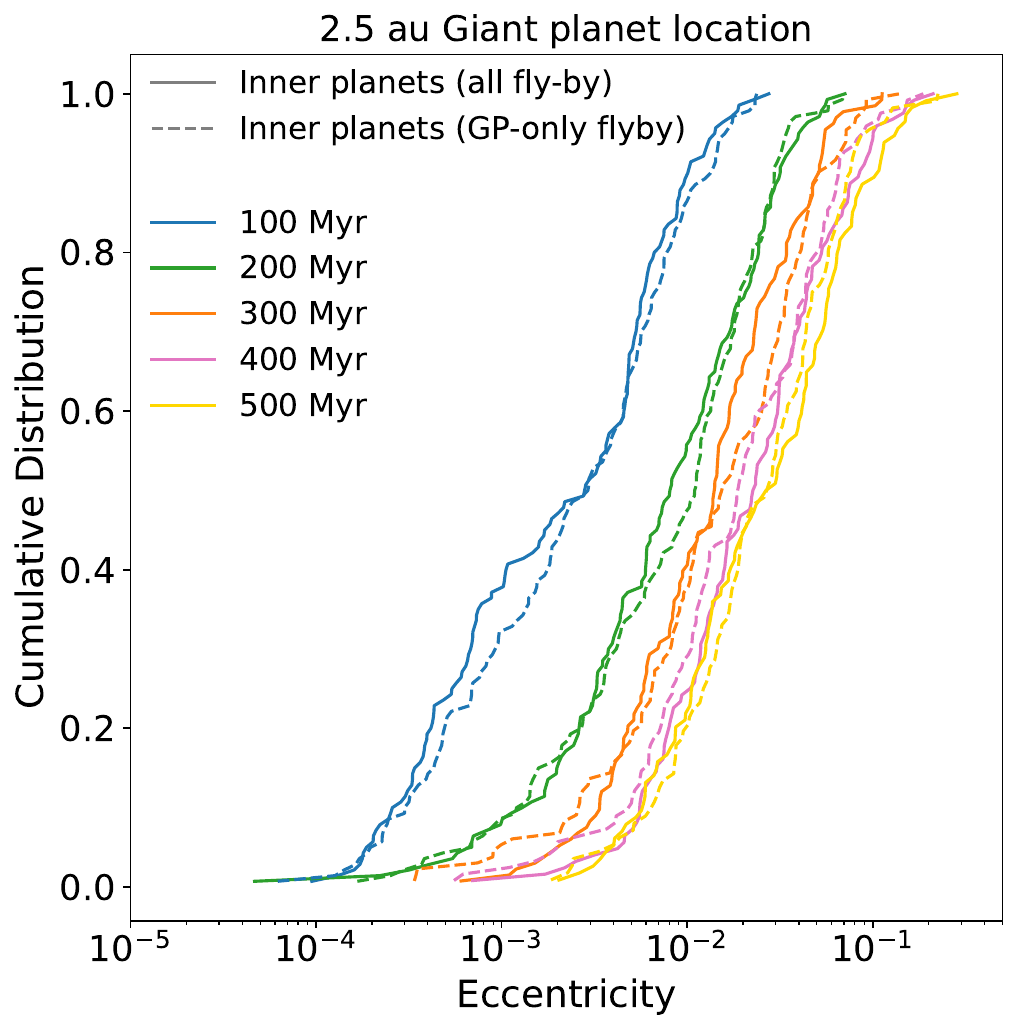}
     \end{minipage}
        \centering
        \vspace{0pt}
    \begin{minipage}[t]{0.9\columnwidth}
    	\includegraphics[width=0.95\linewidth]{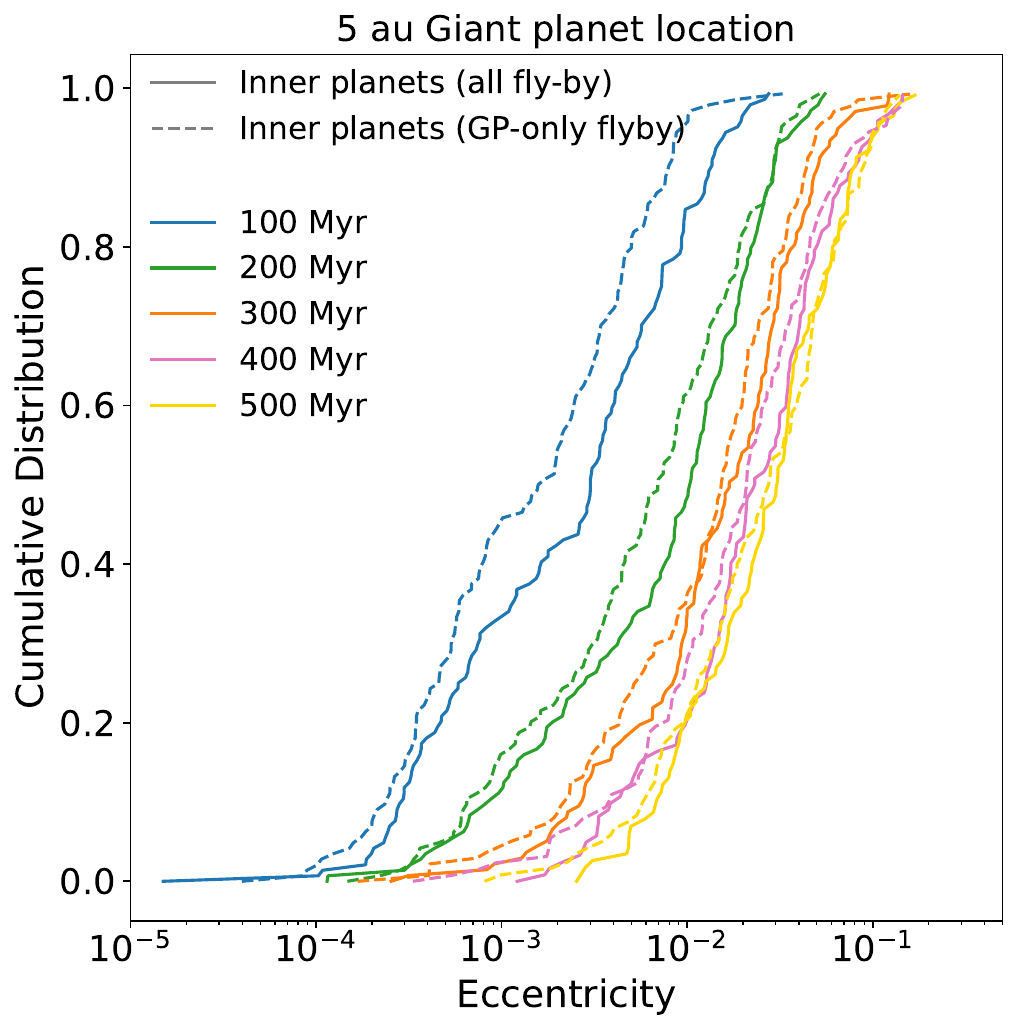}
      \end{minipage}
       \centering
        \vspace{0pt}
    \begin{minipage}[t]{0.9\columnwidth}
         \centering
    	\includegraphics[width=0.95\linewidth]{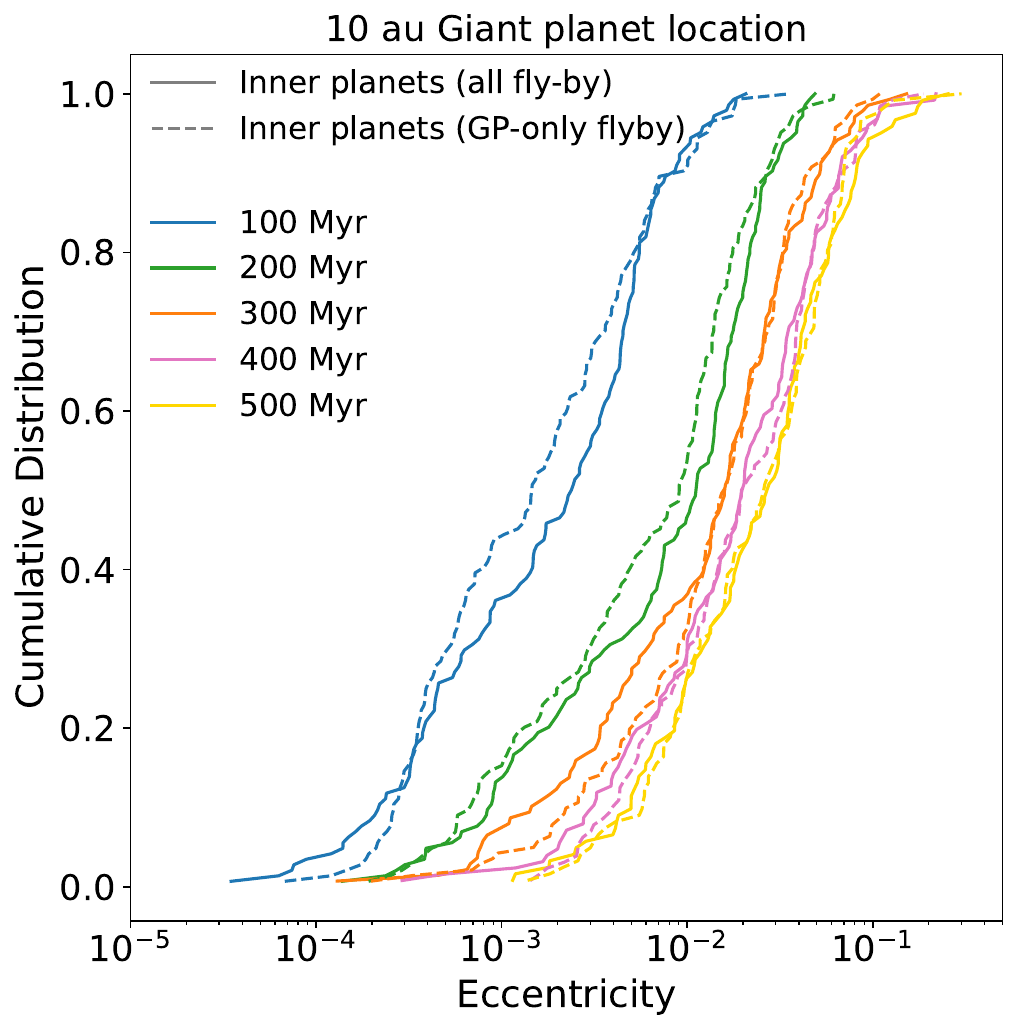}
     \end{minipage}
        \centering
        \vspace{0pt}
    \begin{minipage}[t]{0.9\columnwidth}
    	\includegraphics[width=0.95\linewidth]{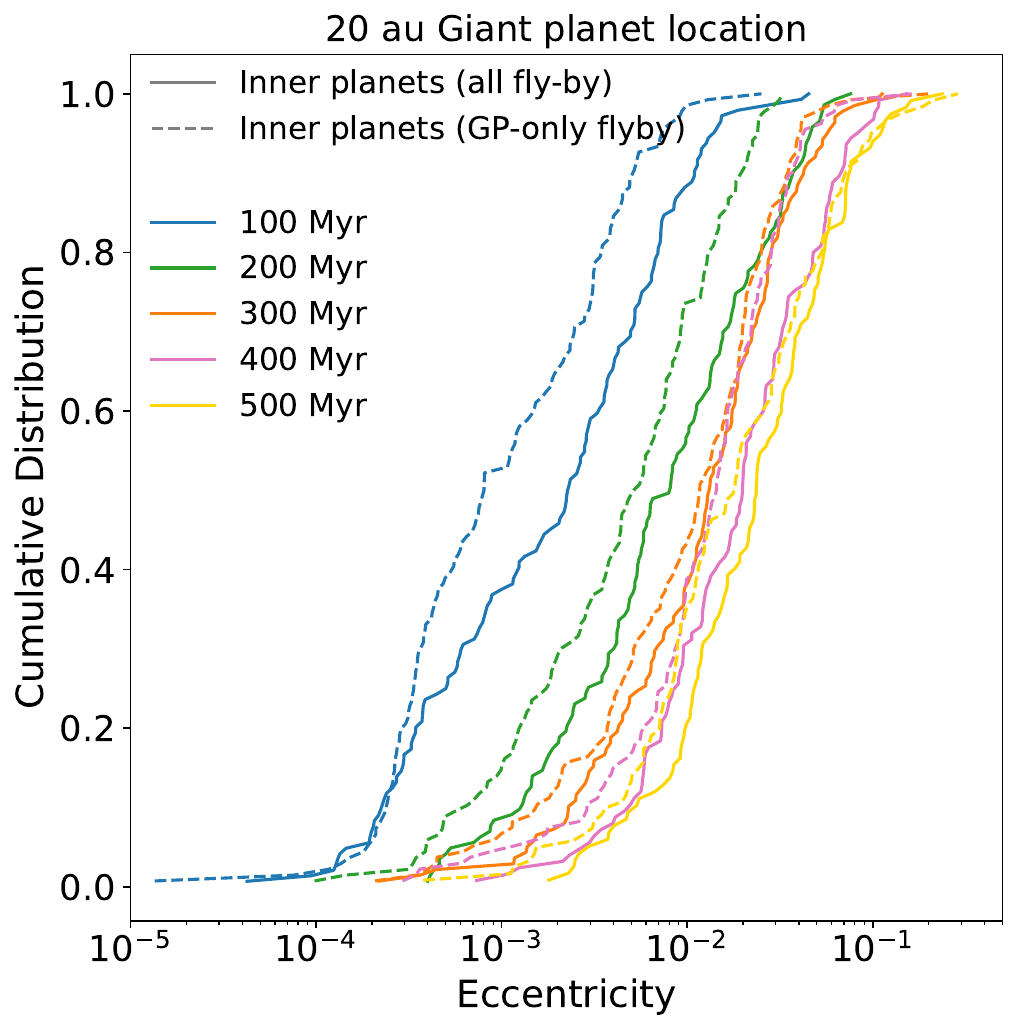}
      \end{minipage}
     \caption{Time evolution of the cumulative distributions of the eccentricity of all remaining inner planets in simulations with initial separation of 14 R$_{MH}$ for different placements of the giant planet. We compare the eccentricity for the close-in planets subjected to a fly-by ('solid` lines) and those where only the distant giant was subjected to that fly-by ('dashed` lines). }
     \label{fig:Ecc_evolv_14}
\end{figure*}

\subsection{The impact of the mass of the giant planet}

Our set of simulations with a different giant planet mass establish the role this difference has on the dynamical evolution of the inner planets. The dashed lines in Fig. \ref{fig:Inc_ecc_GP_only_10_14_lowmass} show the changes in the orbital parameters for the lower-mass giant planet at 1 Myr in comparison to those for the higher-mass giant (solid lines). The eccentricity distributions for the two different masses are very similar for the \textcolor{changes}{2.5, }5, 10 and 20 au locations, with the same increase in eccentricity for the top third of the simulations with the 20 au set-up. \textcolor{changes}{The two-sample Anderson-Darling  [AD] test \citep{Pettitt1976ATA} at a significance level of $3\sigma$ shows that none of the visual differences between the different masses in the plot is significant. The cumulative distributions on the right show the change in inclination at 1 Myr compared to the start of the simulations. The shapes of the distributions for both giant masses are very similar, without any significant difference in the AD-test. Both of these plots indicate that giant planets in systems with different masses react in the same way overall to the fly-by, as would be expected since they effectively behave like test particles.}

\section{Perturbation of the close-in planets by fly-by and/or the giant planet}\label{results}

\citet{2024MNRAS.533.3484S} indicated little difference between the perturbing effects of fly-bys on close-in exoplanet systems with giant planets at different locations after 500 Myr of evolution time. They found no significant differences between the evolution of the average number of remaining inner planets for different giant planet locations (and even in the complete absence of the giant planet). Fig. \ref{fig:Ecc_evolv_10_14_all_comp} uses many of the same simulations, but not those that have a different starting point for the inner planets, as it was shown not to impact the level of perturbation. In this figure, we compare the eccentricities of the remaining close-in planets and find that there is no significant difference in the distributions without or with a giant planet at different locations. This is true for both initial separations for the close-in planets (10 R$_{MH}$ - left - and 14 R$_{MH}$ - right). 

We now compare the eccentricity after external (all planets are subjected to a fly-by) and internal (only the giant planet is subjected to a fly-by) perturbations, and then focus on the evolution of the number of inner planets over time.

\subsection{Evolution of eccentricity of the close-in planets}

In Figs. \ref{fig:Ecc_evolv_10} (10 R$_{MH}$ initial separation) and \ref{fig:Ecc_evolv_14} (14 R$_{MH}$ initial separation), we show the cumulative distributions of the time evolution of the eccentricities split by whether all planets were subjected to the fly-by (solid lines) compared to the case when only the distant giant was perturbed by the fly-by (dashed lines) and then subsequently perturbed a pristine inner planet system. 

For the 10 R$_{MH}$ initial separation planets in Fig. \ref{fig:Ecc_evolv_10}, the distributions show little visual differences for the 2.5 and 5 au giant planet location (top row) distributions over the 500 Myr simulation time. Like before, we use a \textcolor{changes}{two-sample AD-test to establish statistical significance at a $3\sigma$ level. The 2.5 and 5 au distributions are not significantly different at any of the tested times during the 500 Myr simulation time. The 10 au distributions (lower left) diverge significantly between the internal and external case for the initial $\sim$200 Myr, but at 300, 400 and 500 Myr, they are similar. When only the giant planet experiences the fly-by in the 20 au scenario (bottom right), where it is positioned farthest from the inner planetary system, the close-in planets maintain lower eccentricities and are significantly different throughout the 500 Myr simulation time. They start similarly different as in the 10 au case. But instead of developing more similar eccentricities throughout the simulation time, the giant planet fly-by-only eccentricities remain significantly smaller than the all-planet fly-by ones and show different curves at 500 Myr.}

\begin{figure*}
    \centering
         \centering
    	\includegraphics[width=0.9\linewidth]{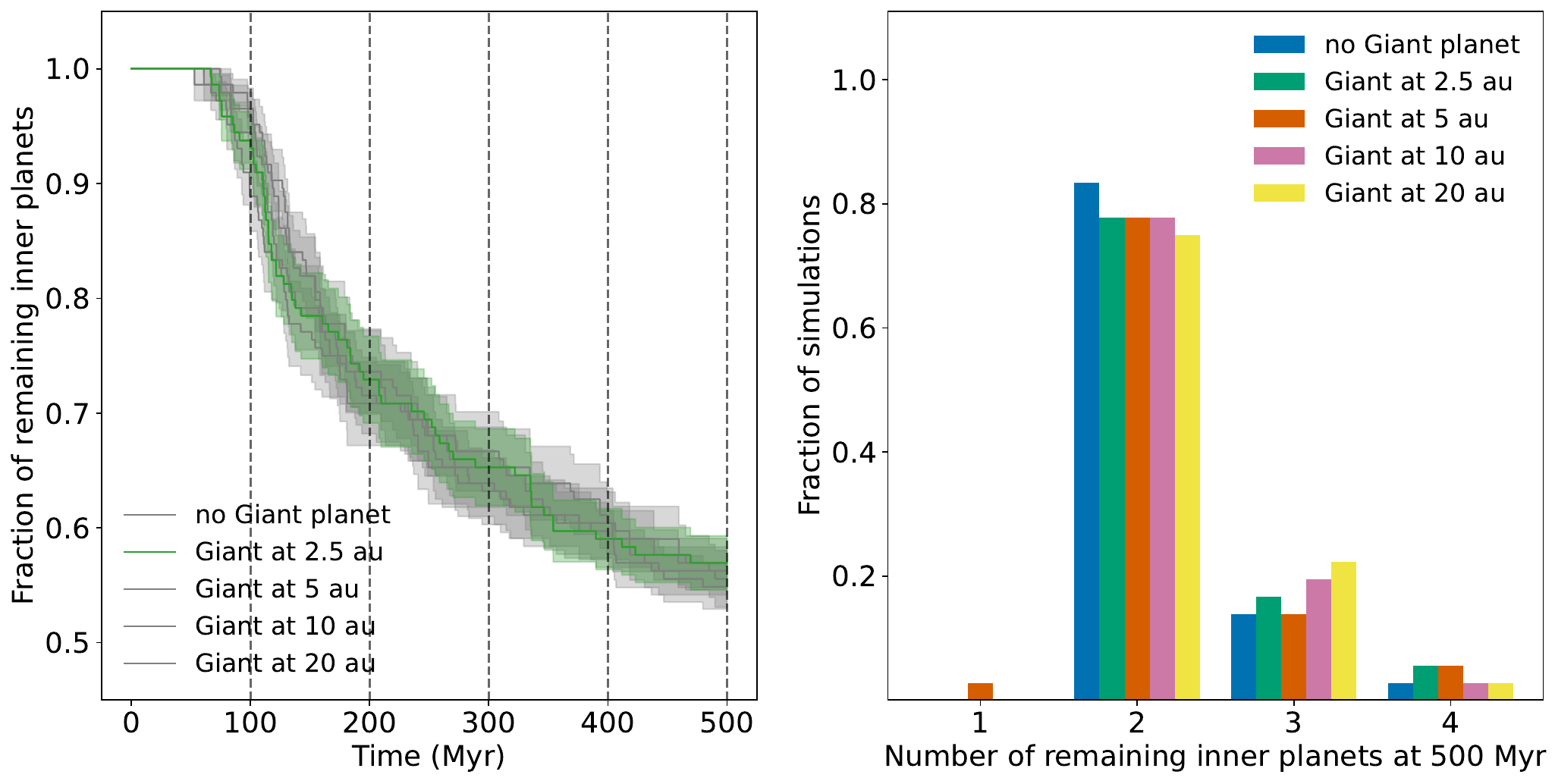}
        \centering
        \vspace{0pt}
     \label{fig:Num_plan_evolve_10_all_GP}
    \centering
         \centering
    	\includegraphics[width=0.9\linewidth]{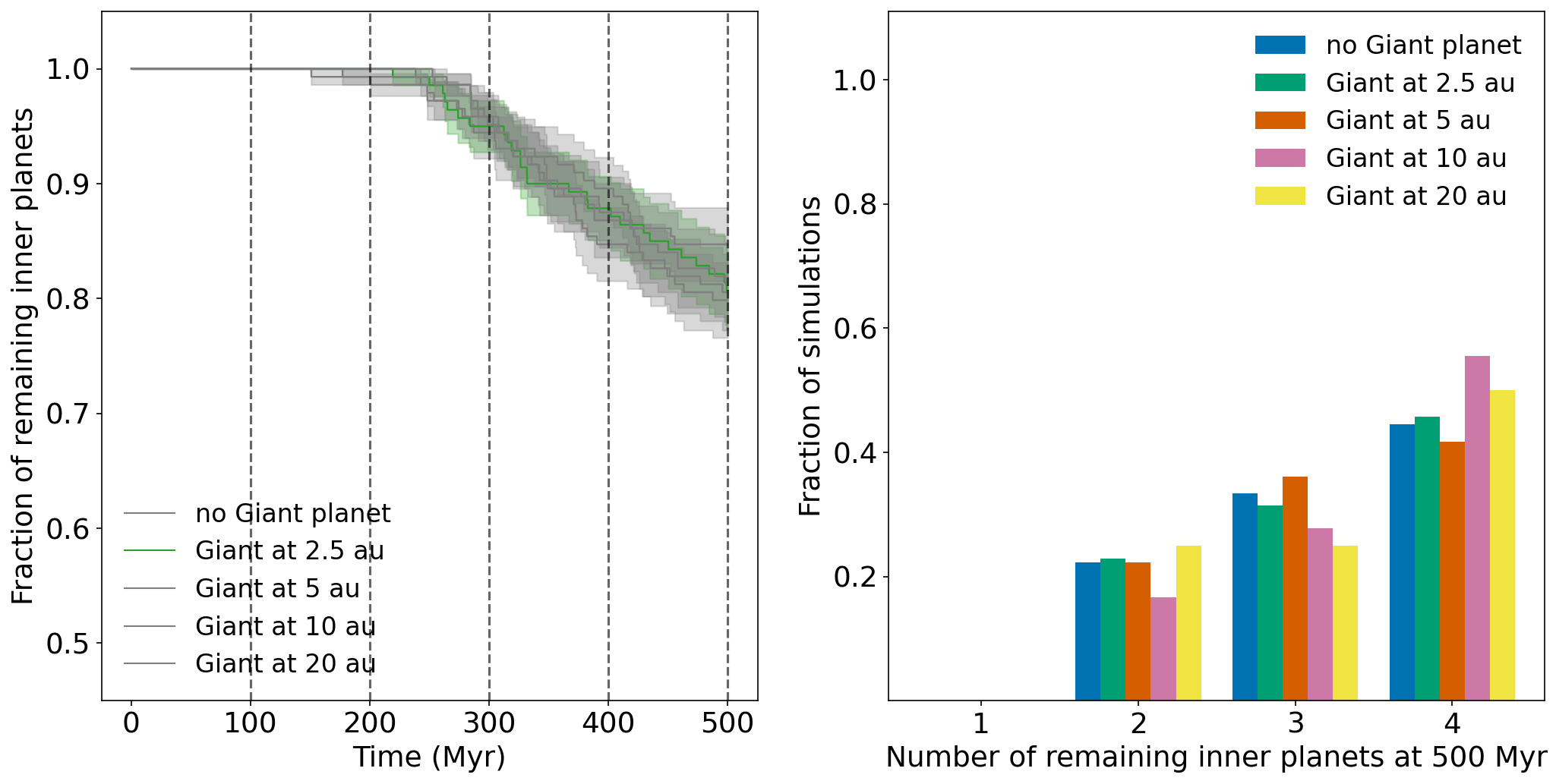}
        \centering
        \vspace{0pt}
   \caption{\textbf{Top row: 10 R$_{\rm{MH}}$ - Left:} This inner planet number evolution plot uses the same data as Fig. 3 from \citet{2024MNRAS.533.3484S} for the close-in planet system with a closer initial separation. We add data for the closer location of 2.5 au and show that even with this closer placement of the giant planet, the reaction of the close-in planet system to the different fly-bys does not change. These systems move from a four-planet to a two-planet inner configuration. \textbf{Right:} These histograms show the number of close-in planets at 500 Myr with the same fly-by scenarios as on the left. We find that our initial four-planet inner systems predominantly turn into two-planet systems after 500 Myr. \textbf{Bottom row: 14 R$_{\rm{MH}}$ - Left:} This inner planet number evolution plot uses the same data as Fig. 4 from \citet{2024MNRAS.533.3484S}for the close-in planet system with an initially further inner planet separation. We add data for the closer location of 2.5 au and show that even with this closer placement of the giant planet, the reaction of the close-in planet system to the different fly-bys does not change. These systems stay as four-planet systems for much longer than for the closer initial separation setup. \textbf{Right:} These histograms show the number of close-in planets at 500 Myr with the same fly-by scenarios as on the left. We find that our initial four-planet inner systems gradually move towards two-planet inner systems, but overall, for fewer of the simulations. We expect this to be a delayed time effect and that these systems will move towards fewer planets but over a longer time frame.}
     \label{fig:Num_plan_evolve_10_14_all_GP}
\end{figure*}

For the 14 R$_{MH}$ initial separation planet systems in Fig. \ref{fig:Ecc_evolv_14}, the close-in planets are initially spaced further apart than in the previous case. When comparing their eccentricity evolution for the different giant planet locations, we find similar results to those for the closer initial mutual separation case. The 2.5 au giant planet location simulations show a similar eccentricity distribution over their simulation time, i.e. the \textcolor{changes}{two-sample AD-test does not show any significant difference. For the 5 au location, the all-fly-by simulations show a small visual tendency towards higher eccentricities for the remaining inner planets initially at 100 Myr. Still, this difference is not significant and disappears for all later times. Unlike the closer initial separation planet simulations, for the 10 au simulations with 14 R$_{MH}$ separation, the eccentricity is similar between the internal and external cases at all times at the $3\sigma$ significance level. Finally, for the 20 au placement of the giant planet, the eccentricities for the giant-only fly-by simulations show visually smaller values throughout the simulation time, but the curves are statistically indistinguishable at the 3$\sigma$ level throughout the 500 Myr.}

\begin{figure*}
    \centering
         \centering
    	\includegraphics[width=0.9\linewidth]{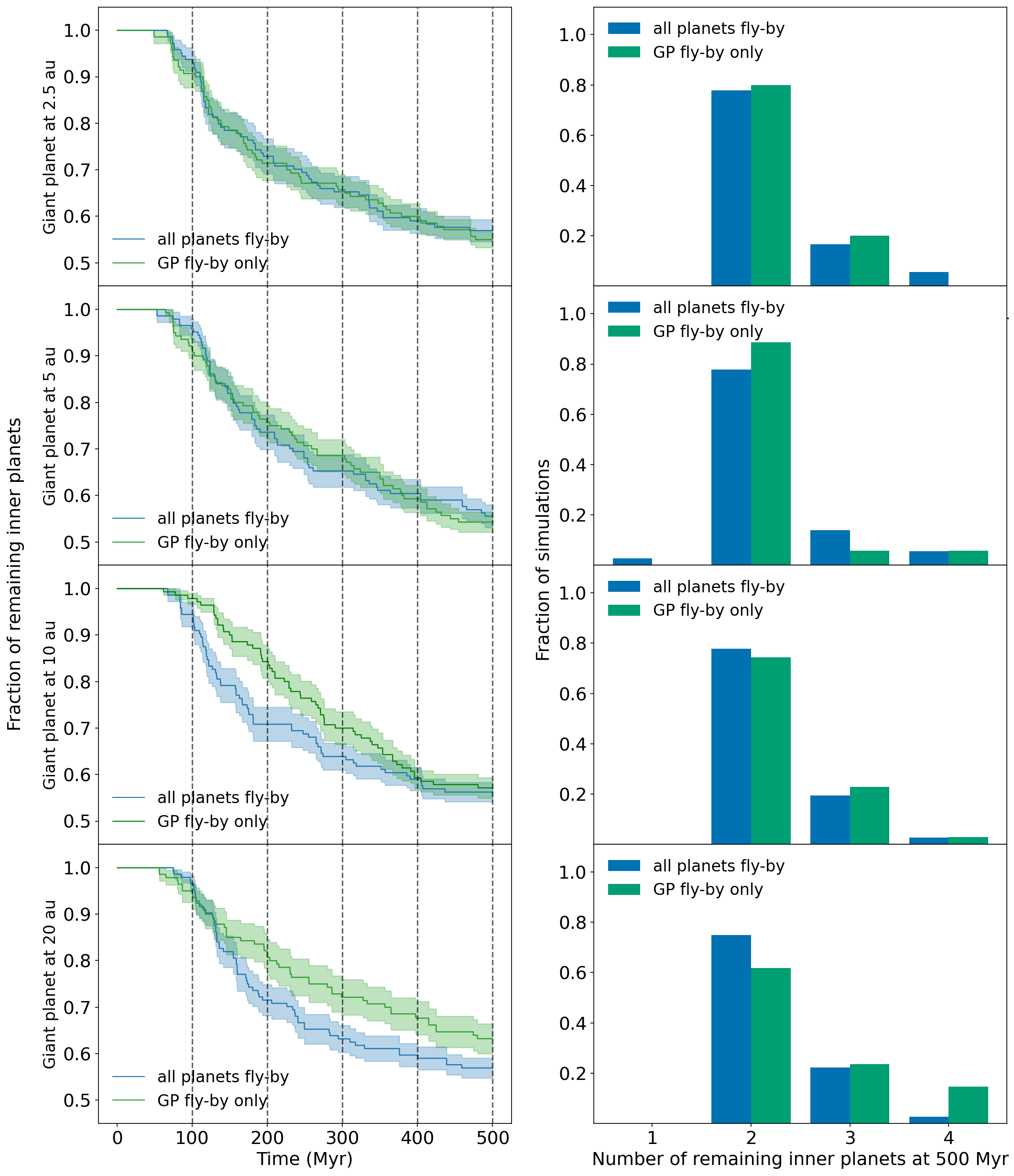}
        \centering
     \caption{\textbf{Left:} The evolution of the average number of close-in planets is shown over 500 Myr (always starting with 4 planets) for simulations where these planets are initially placed at a separation of 10 R$_{\rm{MH}}$. From top to bottom row, we show the differences in the evolution depending on the position of the giant planet and compare the evolution of ''all 5 planets`` and ''GP only`` subjected to a fly-by. The shaded regions depict the standard error of the mean [SEM] as a measure of significance. We find that for three of the four giant locations, there is no significant difference between the average number of these planets in these two fly-by scenarios. The 20 au giant placement shows a slightly slower planet number evolution in the ''GP only`` fly-by scenario, but we suspect that this is merely a delay. \textbf{Right:} These histograms show the number of close-in planets at 500 Myr with the same differences in the fly-by scenarios as on the left. We find that our initial four-planet inner systems predominantly turn into two-planet systems after 500 Myr.}
     \label{fig:Num_plan_evolve_10_all_comp}
\end{figure*}

\begin{figure*}
    \centering
         \centering
    	\includegraphics[width=0.9\linewidth]{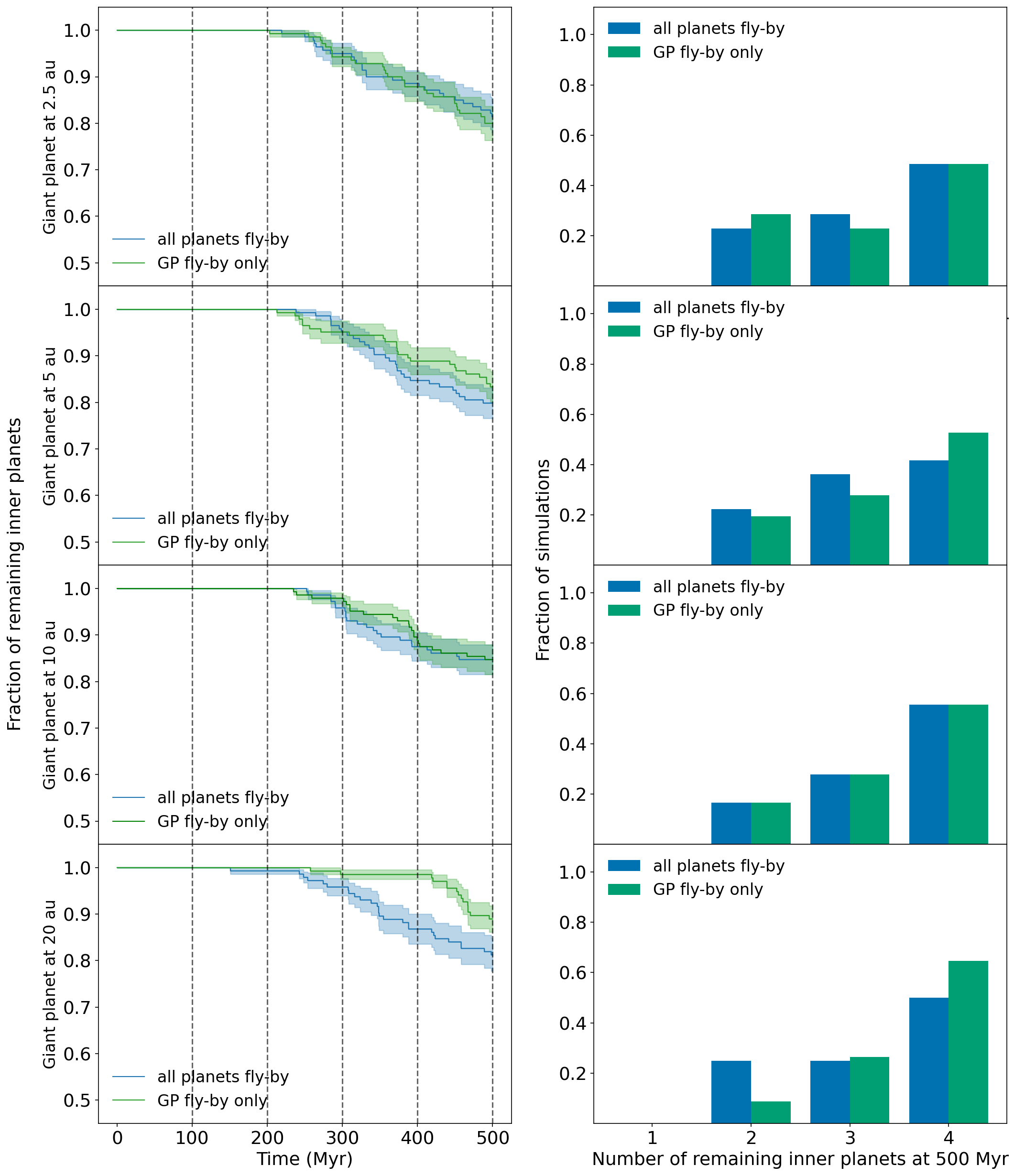}
        \centering
        \vspace{0pt}
   \caption{\textbf{Left:} The evolution of the average number of close-in planets is shown over 500 Myr (always starting with 4 planets) for simulations where these planets are initially placed at a separation of 14 R$_{\rm{MH}}$. From top to bottom row, we show the differences in the evolution depending on the position of the giant planet and compare the evolution of ''all 5 planets`` and ''GP only`` subjected to a fly-by. We find that for three of the four giant locations, there is no significant difference between the average number of these planets in these two fly-by scenarios. Once again, the 20 au giant placement shows a slightly slower planet number evolution in the ''GP only`` fly-by scenario, but we suspect that this is merely a delay. \textbf{Right:} These histograms show the number of close-in planets at 500 Myr with the same differences in the fly-by scenarios as on the left. We find that our initial four-planet inner systems predominantly turn into two-planet systems after 500 Myr.}
     \label{fig:Num_plan_evolve_14_all_comp}
\end{figure*}

\subsection{Evolution of number of close-in planets}\label{res_transits}

We now evaluate the evolution of the average number of close-in planets over the 500 Myr simulation time. In Fig. \ref{fig:Num_plan_evolve_10_14_all_GP}, we show the evolution of the average number of inner planets for systems where all planets were subjected to a fly-by. From \citet{2024MNRAS.533.3484S}, we know that there is no significant difference between the different giant planet locations and even a system without a giant planet shows a similar evolution pattern of inner-planet numbers. In Fig. \ref{fig:Num_plan_evolve_10_14_all_GP}, we now add the 2.5 au location, which is closest to the inner planets, but furthest away from the perturbing effect of the fly-by star. On the left in the figures for both initial separations, we see that the 2.5 au giant planet shows the same evolution behaviour of the number of close-in planets as all other giant locations. On the right, the histograms show the number of remaining inner planets at 500 Myr, and we find that the new giant planet location slots in with the other locations. The four initially closer inner planet separation systems (top row - 10 R$_{\rm{MH}}$) predominantly evolve into two-planet systems at the end of our simulations. Whereas, the initial further planet placements show a mix of four, three and two-inner planet systems. Running these simulations for longer would be required to see whether they reached a general two-planet configuration for a large proportion of systems. 

Next, we compare the effects of internal versus external perturbations by evaluating the evolution of the number of inner planets for each of the giant planet locations. In Fig. \ref{fig:Num_plan_evolve_10_all_comp}, we show the 10 R$_{\rm{MH}}$ initial separation systems, from top to bottom, with 2.5, 5, 10 and 20 au giant planet locations. We find that for the 2.5 and 5 au locations, there is no visual difference between the evolution of the number of planets for internally or externally perturbed systems. The 10 au giant planet location simulations start exhibiting collisions, i.e. a reduction in the number of planets around the same time, regardless of internal or external perturbations. But the inner planet systems that are externally perturbed (all planets experiencing fly-by) collide more frequently, and the fraction of remaining inner planets drops rapidly after the onset of the first collisions. The internally perturbed inner planets (by an eccentric/inclined giant planet only) have a more gradual reduction of the number of inner planets. Regardless of the steepness of their planet number evolution, both of these curves arrive at the same fraction of inner planets ($\sim$0.6) around the 400 Myr mark and continue evolving in the same flat pattern up to 500 Myr. The histogram shows a similar distribution of inner planet numbers at 500 Myr despite the differences in preceding evolution. Finally, the 20 au simulations start with collisions for the two cases around the same time and show a similar evolution for the first $\sim$50 Myr. At around 130 Myr, the evolution diverges, with the fraction of remaining inner planets dropping much more quickly for the external perturbation case compared to the internal perturbation case. After around 200 Myr, these two curves then evolve in parallel until the end of our simulations at 500 Myr and remain different in the mean values. We now analyse the average number of planets at 500 Myr, i.e. the data in the histograms on the right, with the \textcolor{changes}{Chi-Squared [$\chi^2$] Test \citep{pearson1900x} at a significance level of $3\sigma$.} The 2.5 and 5 au planet number distributions are not significantly different at 500 Myr, nor is the 10 au set. For the 20 au case, the evaluation based on the SEM (standard error of the mean) indicates a possible difference at that time due to a non-overlap. However, the distribution of individual planet numbers is not significantly different as measured by the \textcolor{changes}{$\chi^2$-test} at the end of our simulations at a $3\sigma$ level.

In Fig. \ref{fig:Num_plan_evolve_14_all_comp}, we show the evolution of the number of remaining planets for the 14 R$_{\rm{MH}}$ initial separation systems. When compared to the initially closer placement of the inner planets, we find that these systems start to show collisions of their inner planets due to overlapping orbits much later, which is not surprising due to their further apart placement. When comparing the effect of external versus internal perturbations, we find that the 2.5, 5 and 10 au giant planet location simulations show a similar evolution. Collisions start around the same time, and the steepness of the evolution curves is similar. The histograms on the right show the number of remaining planets at 500 Myr, and while there are differences in the systems showing different numbers of planets, the overall trend is similar for the three closer giant planet placements. The \textcolor{changes}{$\chi^2$-test} for these three giant planet placements shows that the internally and externally perturbed evolutions of the planet numbers are not significantly different. Like before, the 20 au giant planet placement systems' mean values suggest a significant difference in the evolution of their mean fraction of planets, as indicated by the SEM non-overlap. The externally perturbed systems start colliding first at around 150 Myr, but then take a further $\sim$100 Myr to show collisions in a larger number of simulations. The internally perturbed systems show the first collisions much later ($\sim$70--80 Myr), and only in a few of our simulations. The drop in fraction becomes more noticeable only after 400 Myr, and at 500 Myr, the internally perturbed systems have a significantly larger number of simulations still showing four inner planets. This difference is also evident in the histograms for the 20 au final state at 500 Myr. For the internally perturbed systems, the number of two-planet inner systems differs considerably from those systems still in their original four-planet state. As in the previous case with 10 R$_{\rm{MH}}$ initial separation, the mean numbers are different, as shown by the non-overlapping SEM, but the underlying individual planet number distributions at 500 Myr are not significantly different at a $3\sigma$ level \textcolor{changes}{in the $\chi^2$-test.}
\\
\\
\begin{figure}
    \centering
    \begin{minipage}[t]{0.9\columnwidth}
         \centering
    	\includegraphics[width=1.0\linewidth]{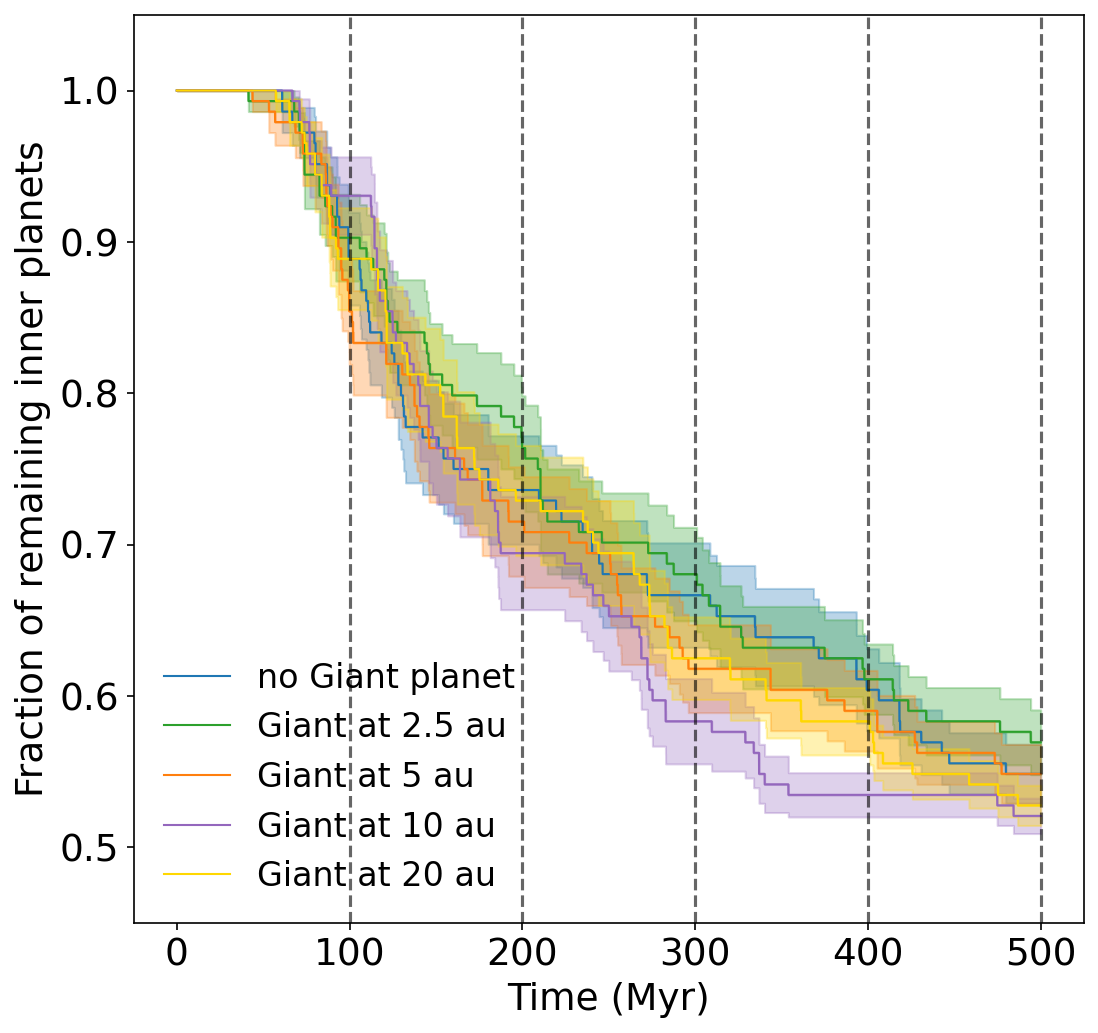}
     \end{minipage}
        \centering
    \begin{minipage}[t]{0.9\columnwidth}
    	\includegraphics[width=1.0\linewidth]{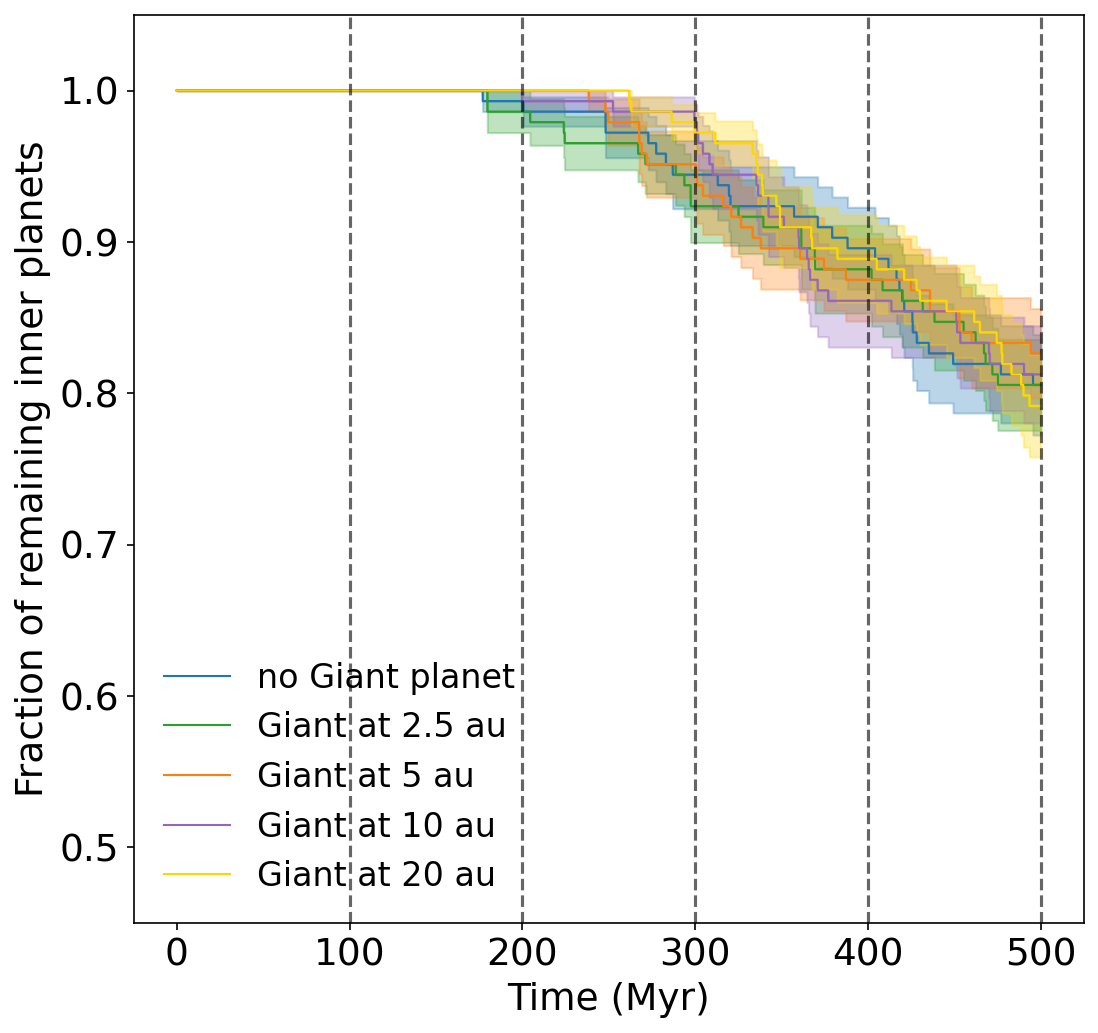}
      \end{minipage}
     \caption{These two plots (top: 10 R$_{\rm{MH}}$ separation, bottom: 14 R$_{\rm{MH}}$ separation) show the comparison between the externally perturbed inner-planet systems for all giant planet distances with a lower giant planet mass. Also plotted is the evolution of the systems where the Giant planet is completely absent. We find that, as in the higher giant mass case, there is no significant difference between the evolution of the average planet numbers, regardless of distance and the absence/presence of the giant. As the no giant curve is the same as in the top (10 R$_{\rm{MH}}$) and bottom graph (14 R$_{\rm{MH}}$) in Fig. \ref{fig:Num_plan_evolve_10_14_all_GP}, this means that the giant planet mass has no influence on the subsequent evolution of the inner planet number in the externally perturbed scenario.}\label{fig:cdf_plan_num_lowmass_10_ext}
\end{figure}

\noindent\textbf{The effect of a change of mass of the giant planet}

In the simulation where we reduce the mass of the giant planet from 5 M$_{\rm{Jup}}$ to 1 M$_{\rm{Jup}}$, we find that the evolution of the number of planets does not differ from the evolution of the planet systems without a giant planet, as illustrated in Figs. \ref{fig:cdf_plan_num_lowmass_10_ext} (top and bottom show the different initial planet separations, with histograms showing the status at 500 Myr available in Appendix,  Fig. \ref{fig:low_mass_hist}). As such, the lower-\textcolor{changes}{mass }planet simulations subjected to a fly-by also show a similar planet number evolution as the fly-by simulations with a higher-mass planet.

When comparing the all-planet fly-bys to the giant planet-only fly-bys, we find that despite the giants (lower or higher mass) being similarly perturbed by the fly-by, the lower-mass giant influences the close-in planets much less than the higher-mass giant. Over the 500 Myr simulation time, the perturbed lower-mass giant planets do not exert the same influence as the higher-mass ones and do not result in any collisions over our covered simulation time (see Appendix Figs. \ref{fig:plan_num_lowmass_10_int} and \ref{fig:plan_num_lowmass_14_int}). 

\section{Discussion}\label{discussion}
This current study expands on our previous work in \citet{2024MNRAS.533.3484S}, where we investigated the reaction of a close-in planet system to fly-by encounters that are common in young star-forming regions. We showed that fly-bys can lead to a decrease in the number of inner planets as well as an increase in mutual inclinations, leading to these systems no longer appearing as multi-planet transiting systems. In this current analysis, we follow up on the finding that the inner planet systems react similarly to the fly-bys regardless of the presence/absence of a giant planet and the exact location of that giant. We investigate if the inner planets behave differently in the absence of a fly-by, where only the giant planet was perturbed by the close encounter (internal perturbation) and compare these to the situation of all planets having been subjected to the fly-bys (external perturbation). We also vary the mass of the giant planet from 5 M$_{\rm{Jup}}$ to 1 M$_{\rm{Jup}}$ to evaluate the effect of the mass. 

With regard to the perturbation of the giant planet, we find that this planet departs from its initial circular, coplanar architecture after being perturbed by the fly-by, which we take as the starting point for the internal perturbation investigation. The effect on the giant is particularly apparent for the 20 au planet locations, as these are closest to the perturbing stars. The inclination changes for the giant planets get higher the further out the giant planet is located. When the giant planet is located at the inner locations of 2.5 and 5 au, the fly-by timescale is larger than the giant planet period in all cases. These encounters are adiabatic and have a smaller effect on the orbits. For the further out locations at 10 au and 20 au, the encounters start to impart impulsive kicks on the giant planet when the planet's period is larger than the fly-by timescale \citep[e.g.][]{2003gmbp.book.....H,2009ApJ...697..458S, 2023MNRAS.526.1987F}. This results in a quicker evolution of the giant planet's orbit away from its initial circular and coplanar state. The immediate reaction of the giant planet to the fly-bys differs very little based on the mass of the giant planet, with similar levels of eccentricity and change in inclination. This, therefore, leaves a similar starting position for the internal perturbation evaluation. 

The reaction of the giant planet to the fly-bys in our simulations is no different to what other authors found in their analyses of these interactions. \citet{2022MNRAS.514..920D} showed that encounters in the birth cluster can alter the orbits of giant planets to be more eccentric and inclined. \citet{2009ApJ...697..458S} showed that the closer the encounter distance, the larger the change in the orbital parameters of the planets involved, similar to our results. An example of the effect of perturbations on giant planets on orbits wider than what we have chosen is presented in \citet{2023MNRAS.525.1912C}. The authors found that a single encounter suffices to perturb giant planets significantly to show altered orbits. Other works showed similar results for perturbed orbits after stellar encounters \citep[e.g.][]{2006ApJ...640.1086F,2011MNRAS.411..859M,2012MNRAS.419.2448P,2018MNRAS.474.5114C}.

After establishing that the fly-by perturbs the giant planet, making it more eccentric and inclined (regardless of the giant's mass), we can evaluate how much effect this planet has on internally perturbing the inner coplanar, circular planet system and how this compares to the external perturbation. Our analysis in this work shows that the eccentricity distributions (as a proxy for orbital alterations) of the close-in planets do not seem to be majorly driven by the kind of perturbation (internal or external), but rather develop similarly. The externally perturbed systems include additional internal perturbations from the giant planets. Still, the simulations lacking a giant planet altogether highlight that this additional perturbation does not affect the outcome in eccentricity and remaining inner-planet numbers. Similarly, the number of inner planets does not differ significantly between the externally perturbed and internally perturbed systems for the placements of the giant planet at 2.5, 5 and 10 au. In addition, we also establish that in the external perturbation set-up, the presence/absence of the giant does not change the outcome of the evolution of the inner planet system. We see a small difference, for the 20 au giant placements for both initial close-in planet separations, in that the internally perturbed systems show significantly more remaining inner planets after 500 Myr. We suspect that this is a delay effect and that, given sufficient additional time, the two curves will converge and show no difference as the other giant planet locations. We prove that this is the case by running all of our 20 au simulations to 600 Myr, with the extended evolution of the planet numbers shown in Appendix Fig. \ref{fig:CDF_20_au_ext}. As suspected, the curves converge after 600 Myr, highlighting a similar end state regardless of the case of perturbation, as the simulations with giants placed at different, closer locations. This highlights that the external direct perturbation is more important for the close-in planetary system evolution in our simulations. For the two giant planet masses studied here (1 and 5 M$_{\rm{Jup}}$), this outweighs the effects from any perturbations on the giants that are then transferred to the inner planets.

Several works have shown that close-in planet systems with a distant eccentric/inclined giant can be perturbed by this outer companion \citep[e.g.][]{2017AJ....153...42L,2017MNRAS.467.1531H,2017MNRAS.469..171R,2017MNRAS.468.3000M,2018MNRAS.478..197P,2019MNRAS.482.4146D,2020MNRAS.498.5166P}, as well as work showing that fly-bys that occur within an evolving star-forming region can cause severe perturbations of multi-planet systems \citep[e.g.][]{2011MNRAS.411..859M,2013MNRAS.433..867H,2017MNRAS.470.4337C,2019MNRAS.489.4311C,2019A&A...624A.120V,2020MNRAS.497.1807S,2022MNRAS.512.2460S,2022MNRAS.509.5253W,2023MNRAS.520..637R}. There is little other work comparing the perturbing effects of fly-bys on close-in planet systems with a distant giant planet, separating the effect of the perturbed giant and the fly-by itself. \citet{2019AJ....158...94B} investigated the interplay between internal and external perturbations on exoplanets on significantly wider orbits than ours in this work. Their work concludes that a fly-by can interrupt planet-planet scattering, effectively ending this internal perturbation process. Their planets were put on orbits that, based on \citet[][]{2015ApJ...807...44P} would not be stable in isolation, whereas ours are stable up to at least 500 Myr. These systems, therefore, would exhibit planet-planet scattering in the absence of an external perturber. On the other hand, our exoplanet systems are stable in the absence of an external influence; therefore, the fly-by causes either the destabilisation of the giant alone or of all planets in the system. \citet{2020CeMDA.132...12S} evaluated the influence of internal and external effects on Trans-Neptunian Objects. They concluded that the influence of externally passing stars causes occasional jumps of the objects in the parameter space, but that a close encounter is required. \citet{2011MNRAS.411..859M} showed that a fly-by can leave a planet system in an unstable state after perturbations of the orbits eventually causing planet-planet scattering over millions to hundred million-year timescales. In an analysis by \citet{2022MNRAS.515.5942B}, they showed that a fly-by can induce small perturbations to a simulated planet system, which is not enough to destabilise it immediately. Over time though, this can be transferred between the planets, leading to stronger perturbations.

The results of the work presented here suggest that planet systems like the ones we use do not ``care'' how they are perturbed, but that they are perturbed at all. Their dynamical evolution is affected by any source of perturbation in a similar way and appears to lead to a preferred, stable state, e.g. the 10 R$_{\rm{MH}}$ simulations show that our close-in planets have a preferred state as two-planet systems after 500 Myr starting from four planets. The initially more separated 14 R$_{\rm{MH}}$ simulations show the same general trend of losing planets to inner planet - inner planet collisions, but still have a slight preference for their original four-planet state at 500 Myr.  We base this on the fact that the 10 R$_{\rm{MH}}$ showed this planet number distribution at an earlier time in their evolution.

\textcolor{changes}{
\subsection{Tests of Numerical Approach}
At first thought our results are surprising: that a distant fly-by (albeit close in the context of realistic expected fly-bys), can impact the close-in exoplanets' dynamical evolution. Therefore, it is important to rule out possible causes of numerical or statistical errors. For example, we only chose to start with a single initial configuration (which we confirmed is not close to mean-motion resonances). This was in part based on results from \citet{2024MNRAS.533.3484S}, which found statistically indistinguishable results independent of the planet; however, without a large-scale exploration, something that is computationally infeasible for our run-times to 500 Myr, we cannot rule out that our results are driven by starting from a particular initial configuration. }

\textcolor{changes}{
Furthermore, we remove the perturbing star after 1~Myr in our simulation; at this point the star is at a distance of $\gtrsim 1$~pc, and we switch from the IAS15 integrator to MERCURIUS. We re-ran a subset of our simulations where we modified this procedure, removing the perturber and switching integrator at a range of different (and decoupled) times between 0.8 and 1.2 Myr. We checked energy errors occurring due to our procedure of removing the perturbing fly-by star at the same time as the integrator change. We found that this choice made no difference to the evolution. }

\textcolor{changes}{
We confirmed our choice of MERCURIUS timestep (a percentage of the inner-most planet period) and found that this is generally shorter than one that would have been chosen by the IAS15 integrator for the normal integration of the planetary system without close encounters. During close encounters, the IAS15 integrator takes over and applies an adaptive, short-time timestep \citep{reboundias15}. }

\textcolor{changes}{In principle, a robust numerical test would be to perform our experiment with a different integrator. Running the entire simulation with IAS15 would result in infeasibility; however, one could use the new TRACE algorithm in REBOUND \citep{Lu2024}, or another N-body code. Ultimately, while we have not identified any source of numerical error that is driving our results, with such long integration times ($\sim 10^{10}$ orbits), it is difficult to completely rule them out.}

\textcolor{changes}{Finally, there is a potential impact due to our approach to the reference frame of the simulations. The reference frame is reset to the centre of mass once during all the simulations without the inner planet system reset, which happens at the initial set-up before adding the flyby star. When removing the flyby star in the no-reset simulations, we do not reset the reference frame after the removal of the perturbing star. However, for all the inner reset simulations, we do reset to the centre of mass after the flyby removal when re-inserting the unperturbed inner planets. It is possible this could lead to larger floating-point errors over time as the no-reset systems frame drifts with respect to the barycentre, but would be consistent across all simulations in each of the internal and external sets. These approaches could be investigated in further work on the topic.}

\section{Conclusions}\label{conclusion}

In this paper, we use $N$-body simulations to investigate the effect of internal and external perturbations caused by close encounters that can occur during the early dynamical evolution of a typical young star-forming region on planetary systems containing close-in sub-Neptune or super-Earth planets (with and without a distant giant planet) around a solar-mass host star. This setup is a common exoplanet architecture with occurrence rates of $\gtrsim$30 per cent around FGKM stars \citep[e.g.][]{2018AJ....156...92Z, 2022ApJS..262....1R,2025arXiv250617091D}. We evaluate whether there is a difference between the reactions of the inner planet system (change in orbital parameters and numbers of remaining planets), either to perturbations induced by an inclined/eccentric distant giant planet or by being subjected to a single fly-by encounter.  

\begin{enumerate}
  \item The impact of an external perturbation, like a stellar encounter, on the orbital characteristics of a close-in planet system does not differ in the presence or absence of a giant planet at different locations after 500 Myr. However, there might be small differences during their dynamical evolution.
  \item We find that for two tested giant planet masses at four different locations, the impact of external perturbations on close-in planet systems is the same as measured by their evolution of inner-planet numbers in comparison to the case of an altogether absent giant planet. The addition or location of a giant planet does not change the reaction of the inner planetary system to the perturbations induced by a close stellar encounter, nor does the mass of this giant planet.
  \item When comparing internal and external perturbations, we find that the mass of the giant planet plays a role:
  \begin{itemize}
      \item For the 5 M$_{\rm{Jup}}$ giants, the impact of the internal or external perturbations on the inner-planet system leads to the same evolution of planet numbers as well as orbital parameters, such as eccentricity. For these systems, the origin of the perturbation is of no consequence to how the close-in planets evolve over time. 
      \item For the lower-mass 1 M$_{\rm{Jup}}$ giants, there is a clear difference between the internal and external perturbation case, where the effect of the external case is much stronger than the internal case.
  \end{itemize} 
  \item In summary, we find that the direct impact of the fly-by on the close-in planets is more important than any that arises only from an inclined and/or eccentric giant planet.
\end{enumerate}

\section*{Acknowledgements}

\textcolor{changes}{We are grateful to the referee, Alexander Mustill, for comments and suggestions that improved the manuscript. }This project has received funding from the European Research Council (ERC) under the European Union’s Horizon 2020 Framework Programme (grant agreement no. 853022, PEVAP). JEO is supported by a Royal Society University Research Fellowship. This work was partly supported by STFC grant ST/Y002407/1.

Simulations in this paper made use of the REBOUND N-body code \citep{rebound}. The simulations were integrated using the hybrid symplectic MERCURIUS integrator \citep{reboundmercurius} and IAS15, a 15th order Gauss-Radau integrator \citep{reboundias15}. The SimulationArchive format was used to store fully reproducible simulation data \citep{reboundsa}. 

This work was performed using the DiRAC Data Intensive service at Leicester, operated by the University of Leicester IT Services, which forms part of the STFC DiRAC HPC Facility (www.dirac.ac.uk). The equipment was funded by BEIS capital funding via STFC capital grants ST/K000373/1 and ST/R002363/1 and STFC DiRAC Operations grant ST/R001014/1. DiRAC is part of the National e-Infrastructure. This work was performed using resources provided by the Cambridge Service for Data Driven Discovery (CSD3) operated by the University of Cambridge Research Computing Service (www.csd3.cam.ac.uk), provided by Dell EMC and Intel using Tier-2 funding from the Engineering and Physical Sciences Research Council (capital grant EP/T022159/1), and DiRAC funding from the Science and Technology Facilities Council (www.dirac.ac.uk).

\section*{Data Availability}

The data underlying this article will be shared on reasonable request to the corresponding author.



\bibliographystyle{mnras}
\bibliography{example} 

@ARTICLE{Fulton2021,
       author = {{Fulton}, Benjamin J. and {Rosenthal}, Lee J. and {Hirsch}, Lea A. and {Isaacson}, Howard and {Howard}, Andrew W. and {Dedrick}, Cayla M. and {Sherstyuk}, Ilya A. and {Blunt}, Sarah C. and {Petigura}, Erik A. and {Knutson}, Heather A. and {Behmard}, Aida and {Chontos}, Ashley and {Crepp}, Justin R. and {Crossfield}, Ian J.~M. and {Dalba}, Paul A. and {Fischer}, Debra A. and {Henry}, Gregory W. and {Kane}, Stephen R. and {Kosiarek}, Molly and {Marcy}, Geoffrey W. and {Rubenzahl}, Ryan A. and {Weiss}, Lauren M. and {Wright}, Jason T.},
        title = "{California Legacy Survey. II. Occurrence of Giant Planets beyond the Ice Line}",
      journal = {\apjs},
         year = 2021,
        month = jul,
       volume = {255},
       number = {1},
          eid = {14},
        pages = {14},
          doi = {10.3847/1538-4365/abfcc1},
archivePrefix = {arXiv},
       eprint = {2105.11584},
 primaryClass = {astro-ph.EP},
       adsurl = {https://ui.adsabs.harvard.edu/abs/2021ApJS..255...14F}
}

@ARTICLE{Lu2024,
       author = {{Lu}, Tiger and {Hernandez}, David M. and {Rein}, Hanno},
        title = "{TRACE: a code for time-reversible astrophysical close encounters}",
      journal = {\mnras},
         year = 2024,
        month = sep,
       volume = {533},
       number = {3},
        pages = {3708-3723},
          doi = {10.1093/mnras/stae1982},
archivePrefix = {arXiv},
       eprint = {2405.03800},
 primaryClass = {astro-ph.EP},
       adsurl = {https://ui.adsabs.harvard.edu/abs/2024MNRAS.533.3708L}
}

@article{pearson1900x,
  title={X. On the criterion that a given system of deviations from the probable in the case of a correlated system of variables is such that it can be reasonably supposed to have arisen from random sampling},
  author={Pearson, Karl},
  journal={The London, Edinburgh, and Dublin Philosophical Magazine and Journal of Science},
  volume={50},
  number={302},
  pages={157--175},
  year={1900},
  publisher={Taylor \& Francis}
}

@ARTICLE{Qiao2023,
       author = {{Qiao}, Lin and {Coleman}, Gavin A.~L. and {Haworth}, Thomas J.},
        title = "{Planet formation via pebble accretion in externally photoevaporating discs}",
      journal = {\mnras},
         year = 2023,
        month = jun,
       volume = {522},
       number = {2},
        pages = {1939-1950},
          doi = {10.1093/mnras/stad944},
archivePrefix = {arXiv},
       eprint = {2303.15177},
 primaryClass = {astro-ph.EP},
       adsurl = {https://ui.adsabs.harvard.edu/abs/2023MNRAS.522.1939Q}
}

@ARTICLE{2024A&A...687A.121K,
       author = {{Kong}, Zhihui and {Johansen}, Anders and {Lambrechts}, Michiel and {Jiang}, Jonathan H. and {Zhu}, Zong-Hong},
        title = "{How the presence of a giant planet affects the outcome of terrestrial planet formation simulations}",
      journal = {\aap},
         year = 2024,
        month = jul,
       volume = {687},
          eid = {A121},
        pages = {A121},
              doi = {10.1051/0004-6361/202349043}}

@ARTICLE{2008ApJ...686..621F,
       author = {{Ford}, Eric B. and {Rasio}, Frederic A.},
        title = "{Origins of Eccentric Extrasolar Planets: Testing the Planet-Planet Scattering Model}",
      journal = {\apj},
         year = 2008,
        month = oct,
       volume = {686},
       number = {1},
        pages = {621-636},
          doi = {10.1086/590926}}

@ARTICLE{2017A&A...598A..70S,
       author = {{Sotiriadis}, Sotiris and {Libert}, Anne-Sophie and {Bitsch}, Bertram and {Crida}, Aur{\'e}lien},
        title = "{Highly inclined and eccentric massive planets. II. Planet-planet interactions during the disc phase}",
      journal = {\aap},
         year = 2017,
        month = feb,
       volume = {598},
          eid = {A70},
        pages = {A70},
          doi = {10.1051/0004-6361/201628470}}

@ARTICLE{2019A&A...624A.120V,
       author = {{van Elteren}, A. and {Portegies Zwart}, S. and {Pelupessy}, I. and {Cai}, M.~X. and {McMillan}, S.~L.~W.},
        title = "{Survivability of planetary systems in young and dense star clusters}",
      journal = {\aap},
         year = 2019,
        month = apr,
       volume = {624},
          eid = {A120},
        pages = {A120},
          doi = {10.1051/0004-6361/201834641}}

@ARTICLE{2018MNRAS.478..197P,
       author = {{Pu}, Bonan and {Lai}, Dong},
        title = "{Eccentricities and inclinations of multiplanet systems with external perturbers}",
      journal = {\mnras},
         year = 2018,
        month = jul,
       volume = {478},
       number = {1},
        pages = {197-217},
          doi = {10.1093/mnras/sty1098}}

@ARTICLE{2024MNRAS.528.7202G,
       author = {{Gajendran}, Sridhar and {Jiang}, Ing-Guey and {Yeh}, Li-Chin and {Sariya}, Devesh P.},
        title = "{Constraining planetary formation models using conditional occurrences of various planet types}",
      journal = {\mnras},
         year = 2024,
        month = mar,
       volume = {528},
       number = {4},
        pages = {7202-7210},
          doi = {10.1093/mnras/stae501}}

@ARTICLE{2017MNRAS.468.3000M,
       author = {{Mustill}, Alexander J. and {Davies}, Melvyn B. and {Johansen}, Anders},
        title = "{The effects of external planets on inner systems: multiplicities, inclinations and pathways to eccentric warm Jupiters}",
      journal = {\mnras},
         year = 2017,
        month = jul,
       volume = {468},
       number = {3},
        pages = {3000-3023},
          doi = {10.1093/mnras/stx693}}

@ARTICLE{2020MNRAS.496.1149L,
       author = {{Li}, Daohai and {Mustill}, Alexander J. and {Davies}, Melvyn B.},
        title = "{Flyby encounters between two planetary systems II: exploring the interactions of diverse planetary system architectures}",
      journal = {\mnras},
         year = 2020,
        month = aug,
       volume = {496},
       number = {2},
        pages = {1149-1165},
          doi = {10.1093/mnras/staa1622}}

@ARTICLE{2018AJ....156...92Z,
       author = {{Zhu}, Wei and {Wu}, Yanqin},
        title = "{The Super Earth-Cold Jupiter Relations}",
      journal = {\aj},
         year = 2018,
        month = sep,
       volume = {156},
       number = {3},
          eid = {92},
        pages = {92},
          doi = {10.3847/1538-3881/aad22a}}

@ARTICLE{2019AJ....157...52B,
       author = {{Bryan}, Marta L. and {Knutson}, Heather A. and {Lee}, Eve J. and {Fulton}, B.~J. and {Batygin}, Konstantin and {Ngo}, Henry and {Meshkat}, Tiffany},
        title = "{An Excess of Jupiter Analogs in Super-Earth Systems}",
      journal = {\aj},
         year = 2019,
        month = feb,
       volume = {157},
       number = {2},
          eid = {52},
        pages = {52},
          doi = {10.3847/1538-3881/aaf57f}}

@ARTICLE{2019AJ....158...94B,
       author = {{Bailey}, Nora and {Fabrycky}, Daniel},
        title = "{Stellar Flybys Interrupting Planet-Planet Scattering Generates Oort Planets}",
      journal = {\aj},
         year = 2019,
        month = aug,
       volume = {158},
       number = {2},
          eid = {94},
        pages = {94},
          doi = {10.3847/1538-3881/ab2d2a}}

@ARTICLE{2016ApJ...821...89B,
       author = {{Bryan}, Marta L. and {Knutson}, Heather A. and {Howard}, Andrew W. and {Ngo}, Henry and {Batygin}, Konstantin and {Crepp}, Justin R. and {Fulton}, B.~J. and {Hinkley}, Sasha and {Isaacson}, Howard and {Johnson}, John A. and {Marcy}, Geoffry W. and {Wright}, Jason T.},
        title = "{Statistics of Long Period Gas Giant Planets in Known Planetary Systems}",
      journal = {\apj},
         year = 2016,
        month = apr,
       volume = {821},
       number = {2},
          eid = {89},
        pages = {89},
          doi = {10.3847/0004-637X/821/2/89}}

@ARTICLE{2009ApJ...697..458S,
       author = {{Spurzem}, R. and {Giersz}, M. and {Heggie}, D.~C. and {Lin}, D.~N.~C.},
        title = "{Dynamics of Planetary Systems in Star Clusters}",
      journal = {\apj},
         year = 2009,
        month = may,
       volume = {697},
       number = {1},
        pages = {458-482},
          doi = {10.1088/0004-637X/697/1/458}}

@ARTICLE{2023MNRAS.525.1912C,
       author = {{Carter}, Ethan J. and {Stamatellos}, Dimitris},
        title = "{On the survivability of a population of gas giant planets on wide orbits}",
      journal = {\mnras},
         year = 2023,
        month = oct,
       volume = {525},
       number = {2},
        pages = {1912-1921},
          doi = {10.1093/mnras/stad2314}}

@ARTICLE{2006ApJ...640.1086F,
       author = {{Fregeau}, John M. and {Chatterjee}, Sourav and {Rasio}, Frederic A.},
        title = "{Dynamical Interactions of Planetary Systems in Dense Stellar Environments}",
      journal = {\apj},
         year = 2006,
        month = apr,
       volume = {640},
       number = {2},
        pages = {1086-1098},
          doi = {10.1086/500111}}

@ARTICLE{2015ApJ...807...44P,
       author = {{Pu}, Bonan and {Wu}, Yanqin},
        title = "{Spacing of Kepler Planets: Sculpting by Dynamical Instability}",
      journal = {\apj},
         year = 2015,
        month = jul,
       volume = {807},
       number = {1},
          eid = {44},
        pages = {44},
          doi = {10.1088/0004-637X/807/1/44}}

@ARTICLE{2017MNRAS.469..171R,
       author = {{Read}, Matthew J. and {Wyatt}, Mark C. and {Triaud}, Amaury H.~M.~J.},
        title = "{Transit probabilities in secularly evolving planetary systems}",
      journal = {\mnras},
         year = 2017,
        month = jul,
       volume = {469},
       number = {1},
        pages = {171-192},
          doi = {10.1093/mnras/stx798}}

@ARTICLE{2017AJ....153...42L,
       author = {{Lai}, Dong and {Pu}, Bonan},
        title = "{Hiding Planets behind a Big Friend: Mutual Inclinations of Multi-planet Systems with External Companions}",
      journal = {\aj},
         year = 2017,
        month = jan,
       volume = {153},
       number = {1},
          eid = {42},
        pages = {42},
          doi = {10.3847/1538-3881/153/1/42}}

@ARTICLE{2017MNRAS.467.1531H,
       author = {{Hansen}, Bradley M.~S.},
        title = "{Perturbation of Compact Planetary Systems by Distant Giant Planets}",
      journal = {\mnras},
         year = 2017,
        month = may,
       volume = {467},
       number = {2},
        pages = {1531-1560},
          doi = {10.1093/mnras/stx182}}

@ARTICLE{2018MNRAS.474.5114C,
       author = {{Cai}, Maxwell Xu and {Portegies Zwart}, Simon and {van Elteren}, Arjen},
        title = "{The signatures of the parental cluster on field planetary systems}",
      journal = {\mnras},
         year = 2018,
        month = mar,
       volume = {474},
       number = {4},
        pages = {5114-5121},
          doi = {10.1093/mnras/stx3064}}

@ARTICLE{2021MNRAS.502.3746R,
       author = {{Rodet}, Laetitia and {Lai}, Dong},
        title = "{Inclination dynamics of resonant planets under the influence of an inclined external companion}",
      journal = {\mnras},
         year = 2021,
        month = apr,
       volume = {502},
       number = {3},
        pages = {3746-3760},
          doi = {10.1093/mnras/stab094}}

@ARTICLE{2020CeMDA.132...12S,
       author = {{Saillenfest}, Melaine},
        title = "{Long-term orbital dynamics of trans-Neptunian objects}",
      journal = {Celestial Mechanics and Dynamical Astronomy},
         year = 2020,
        month = feb,
       volume = {132},
       number = {2},
          eid = {12},
        pages = {12},
          doi = {10.1007/s10569-020-9954-9}}

@ARTICLE{2019MNRAS.482.4146D,
       author = {{Denham}, Paul and {Naoz}, Smadar and {Hoang}, Bao-Minh and {Stephan}, Alexander P. and {Farr}, Will M.},
        title = "{Hidden planetary friends: on the stability of two-planet systems in the presence of a distant, inclined companion}",
      journal = {\mnras},
         year = 2019,
        month = jan,
       volume = {482},
       number = {3},
        pages = {4146-4154},
          doi = {10.1093/mnras/sty2830}}

@ARTICLE{2020MNRAS.498.5166P,
       author = {{Poon}, Sanson T.~S. and {Nelson}, Richard P.},
        title = "{On the origin of the eccentricity dichotomy displayed by compact super-Earths: dynamical heating by cold giants}",
      journal = {\mnras},
         year = 2020,
        month = nov,
       volume = {498},
       number = {4},
        pages = {5166-5182},
          doi = {10.1093/mnras/staa2755}}

@ARTICLE{reboundias15,
       author = {{Rein}, Hanno and {Spiegel}, David S.},
        title = "{IAS15: a fast, adaptive, high-order integrator for gravitational dynamics, accurate to machine precision over a billion orbits}",
      journal = {\mnras},
         year = 2015,
        month = jan,
       volume = {446},
       number = {2},
        pages = {1424-1437},
          doi = {10.1093/mnras/stu2164}
}

@ARTICLE{2021A&A...656A..71S,
       author = {{Schlecker}, M. and {Mordasini}, C. and {Emsenhuber}, A. and {Klahr}, H. and {Henning}, Th. and {Burn}, R. and {Alibert}, Y. and {Benz}, W.},
        title = "{The New Generation Planetary Population Synthesis (NGPPS). III. Warm super-Earths and cold Jupiters: a weak occurrence correlation, but with a strong architecture-composition link}",
      journal = {\aap},
         year = 2021,
        month = dec,
       volume = {656},
          eid = {A71},
        pages = {A71},
          doi = {10.1051/0004-6361/202038554}
          }

@ARTICLE{2025arXiv250617091D,
       author = {{Danti}, Claudia and {Lambrechts}, Michiel and {Lorek}, Sebastian},
        title = "{Super-Earth formation in systems with cold giants}",
      journal = {arXiv e-prints},
         year = 2025,
        month = jun,
          eid = {arXiv:2506.17091},
        pages = {arXiv:2506.17091},
          doi = {10.48550/arXiv.2506.17091},
archivePrefix = {arXiv},
       eprint = {2506.17091},
 primaryClass = {astro-ph.EP},
}

@ARTICLE{2022MNRAS.509.1010R,
       author = {{Rodet}, Laetitia and {Lai}, Dong},
        title = "{The impact of stellar clustering on the observed multiplicity of super-earth systems: outside-in cascade of orbital misalignments initiated by stellar flybys}",
      journal = {\mnras},
         year = 2022,
        month = jan,
       volume = {509},
       number = {1},
        pages = {1010-1023},
          doi = {10.1093/mnras/stab3046}
          }

@ARTICLE{rebound,
       author = {{Rein}, H. and {Liu}, S. -F.},
        title = "{REBOUND: an open-source multi-purpose N-body code for collisional dynamics}",
      journal = {\aap},
         year = 2012,
        month = jan,
       volume = {537},
          eid = {A128},
        pages = {A128},
          doi = {10.1051/0004-6361/201118085}
}

@ARTICLE{reboundmercurius,
       author = {{Rein}, Hanno and {Hernandez}, David M. and {Tamayo}, Daniel and
         {Brown}, Garett and {Eckels}, Emily and {Holmes}, Emma and
         {Lau}, Michelle and {Leblanc}, R{'e}jean and {Silburt}, Ari},
        title = "{Hybrid symplectic integrators for planetary dynamics}",
      journal = {\mnras},
         year = 2019,
        month = jun,
       volume = {485},
       number = {4},
        pages = {5490-5497},
          doi = {10.1093/mnras/stz769}
}

@ARTICLE{reboundsa,
       author = {{Rein}, Hanno and {Tamayo}, Daniel},
        title = "{A new paradigm for reproducing and analyzing N-body simulations of planetary systems}",
      journal = {\mnras},
         year = 2017,
        month = may,
       volume = {467},
       number = {2},
        pages = {2377-2383},
          doi = {10.1093/mnras/stx232}
}

@article{RN59,
       author = {{Bressert}, E. and {Bastian}, N. and {Gutermuth}, R. and
         {Megeath}, S.~T. and {Allen}, L. and {Evans}, Neal J., II and
         {Rebull}, L.~M. and {Hatchell}, J. and {Johnstone}, D. and
         {Bourke}, T.~L. and {Cieza}, L.~A. and {Harvey}, P.~M. and {Merin}, B. and
         {Ray}, T.~P. and {Tothill}, N.~F.~H.},
        title = "{The spatial distribution of star formation in the solar neighbourhood: do all stars form in dense clusters?}",
      journal = {\mnras},
         year = "2010",
        month = "Nov",
       volume = {409},
        pages = {L54-L58},
          doi = {10.1111/j.1745-3933.2010.00946.x}}

@article{RN25,
       author = {{Lada}, Charles J. and {Lada}, Elizabeth A.},
        title = "{Embedded Clusters in Molecular Clouds}",
      journal = {\araa},
         year = "2003",
        month = "Jan",
       volume = {41},
        pages = {57-115},
          doi = {10.1146/annurev.astro.41.011802.094844},
}

@article{Pettitt1976ATA,
  title={A two-sample Anderson-Darling rank statistic},
  author={Anthony N. Pettitt},
  journal={Biometrika},
  year={1976},
  volume={63},
  pages={161-168},
  url={https://api.semanticscholar.org/CorpusID:119481227}
}

@ARTICLE{2022ApJS..262....1R,
       author = {{Rosenthal}, Lee J. and {Knutson}, Heather A. and {Chachan}, Yayaati and {Dai}, Fei and {Howard}, Andrew W. and {Fulton}, Benjamin J. and {Chontos}, Ashley and {Crepp}, Justin R. and {Dalba}, Paul A. and {Henry}, Gregory W. and {Kane}, Stephen R. and {Petigura}, Erik A. and {Weiss}, Lauren M. and {Wright}, Jason T.},
        title = "{The California Legacy Survey. III. On the Shoulders of (Some) Giants: The Relationship between Inner Small Planets and Outer Massive Planets}",
      journal = {\apjs},
         year = 2022,
        month = sep,
       volume = {262},
       number = {1},
          eid = {1},
        pages = {1},
          doi = {10.3847/1538-4365/ac7230}}

@ARTICLE{2007ApJ...668.1267F,
       author = {{Fortney}, J.~J. and {Marley}, M.~S. and {Barnes}, J.~W.},
        title = "{Erratum: ``Planetary Radii across Five Orders of Magnitude in Mass and Stellar Insolation: Application to Transits`` (<A href=''/abs/2007ApJ...659.1661F''>ApJ, 659, 1661 [2007]</A>)}",
      journal = {\apj},
         year = 2007,
        month = oct,
       volume = {668},
       number = {2},
        pages = {1267-1267},
          doi = {10.1086/521435}}

@ARTICLE{2007ApJ...659.1661F,
       author = {{Fortney}, J.~J. and {Marley}, M.~S. and {Barnes}, J.~W.},
        title = "{Planetary Radii across Five Orders of Magnitude in Mass and Stellar Insolation: Application to Transits}",
      journal = {\apj},
         year = 2007,
        month = apr,
       volume = {659},
       number = {2},
        pages = {1661-1672},
          doi = {10.1086/512120}}

@ARTICLE{2020RSOS....701271P,
       author = {{Parker}, Richard J.},
        title = "{The birth environment of planetary systems}",
      journal = {Royal Society Open Science},
         year = 2020,
        month = nov,
       volume = {7},
       number = {11},
          eid = {201271},
        pages = {201271},
          doi = {10.1098/rsos.201271}}

@ARTICLE{2015ApJ...809...93D,
       author = {{Dong}, Ruobing and {Zhu}, Zhaohuan and {Whitney}, Barbara},
        title = "{Observational Signatures of Planets in Protoplanetary Disks I. Gaps Opened by Single and Multiple Young Planets in Disks}",
      journal = {\apj},
         year = 2015,
        month = aug,
       volume = {809},
       number = {1},
          eid = {93},
        pages = {93},
          doi = {10.1088/0004-637X/809/1/93}}

@INPROCEEDINGS{2023ASPC..534..685P,
       author = {{Paardekooper}, S. and {Dong}, R. and {Duffell}, P. and {Fung}, J. and {Masset}, F.~S. and {Ogilvie}, G. and {Tanaka}, H.},
        title = "{Planet-Disk Interactions and Orbital Evolution}",
    booktitle = {Astronomical Society of the Pacific Conference Series},
         year = 2023,
       editor = {{Inutsuka}, S. and {Aikawa}, Y. and {Muto}, T. and {Tomida}, K. and {Tamura}, M.},
       series = {Astronomical Society of the Pacific Conference Series},
       volume = {534},
        month = jul,
        pages = {685},
       adsurl = {https://ui.adsabs.harvard.edu/abs/2023ASPC..534..685P}
}

@ARTICLE{2020ApJ...891..166W,
       author = {{Wang}, Shijie and {Kanagawa}, Kazuhiro D. and {Hayashi}, Toshinori and {Suto}, Yasushi},
        title = "{Architecture of Three-planet Systems Predicted from the Observed Protoplanetary Disk of HL Tau}",
      journal = {\apj},
         year = 2020,
        month = mar,
       volume = {891},
       number = {2},
          eid = {166},
        pages = {166},
          doi = {10.3847/1538-4357/ab781b}}

@ARTICLE{2020MNRAS.495.3104S,
       author = {{Schoettler}, Christina and {de Bruijne}, Jos and {Vaher}, Eero and {Parker}, Richard J.},
        title = "{Runaway and walkaway stars from the ONC with Gaia DR2}",
      journal = {\mnras},
         year = 2020,
        month = jul,
       volume = {495},
       number = {3},
        pages = {3104-3123},
          doi = {10.1093/mnras/staa1228}}

@ARTICLE{2022MNRAS.514..920D,
       author = {{Daffern-Powell}, Emma C. and {Parker}, Richard J. and {Quanz}, Sascha P.},
        title = "{The Great Planetary Heist: theft and capture in star-forming regions}",
      journal = {\mnras},
         year = 2022,
        month = jul,
       volume = {514},
       number = {1},
        pages = {920-934},
          doi = {10.1093/mnras/stac1392}}

@ARTICLE{2022MNRAS.515.5942B,
       author = {{Brown}, Garett and {Rein}, Hanno},
        title = "{On the long-term stability of the Solar system in the presence of weak perturbations from stellar flybys}",
      journal = {\mnras},
         year = 2022,
        month = oct,
       volume = {515},
       number = {4},
        pages = {5942-5950},
          doi = {10.1093/mnras/stac1763}}

@ARTICLE{2023MNRAS.520..637R,
       author = {{Rickman}, H. and {Wajer}, P. and {Przy{\l}uski}, R. and {Wi{\'s}niowski}, T. and {Nesvorn{\'y}}, D. and {Morbidelli}, A.},
        title = "{Breakdown of planetary systems in embedded clusters}",
      journal = {\mnras},
         year = 2023,
        month = mar,
       volume = {520},
       number = {1},
        pages = {637-648},
          doi = {10.1093/mnras/stac3705}}

@ARTICLE{2024ApJ...970...97Y,
       author = {{Yu}, Fangyuan and {Lai}, Dong},
        title = "{Free-floating Planets, Survivor Planets, Captured Planets, and Binary Planets from Stellar Flybys}",
      journal = {\apj},
         year = 2024,
        month = jul,
       volume = {970},
       number = {1},
          eid = {97},
        pages = {97},
          doi = {10.3847/1538-4357/ad4f81}}

@ARTICLE{2025A&A...696A.175C,
       author = {{Charalambous}, C. and {Cuello}, N. and {Petrovich}, C.},
        title = "{Breaking long-period resonance chains with stellar flybys}",
      journal = {\aap},
         year = 2025,
        month = apr,
       volume = {696},
          eid = {A175},
        pages = {A175},
          doi = {10.1051/0004-6361/202553710}}

@ARTICLE{2020Natur.586..228S,
       author = {{Segura-Cox}, Dominique M. and {Schmiedeke}, Anika and {Pineda}, Jaime E. and {Stephens}, Ian W. and {Fern{\'a}ndez-L{\'o}pez}, Manuel and {Looney}, Leslie W. and {Caselli}, Paola and {Li}, Zhi-Yun and {Mundy}, Lee G. and {Kwon}, Woojin and {Harris}, Robert J.},
        title = "{Four annular structures in a protostellar disk less than 500,000 years old}",
      journal = {\nat},
         year = 2020,
        month = oct,
       volume = {586},
       number = {7828},
        pages = {228-231},
          doi = {10.1038/s41586-020-2779-6}}

@ARTICLE{2020ApJ...904L...6A,
       author = {{Alves}, Felipe O. and {Cleeves}, L. Ilsedore and {Girart}, Josep M. and {Zhu}, Zhaohuan and {Franco}, Gabriel A.~P. and {Zurlo}, Alice and {Caselli}, Paola},
        title = "{A Case of Simultaneous Star and Planet Formation}",
      journal = {\apjl},
         year = 2020,
        month = nov,
       volume = {904},
       number = {1},
          eid = {L6},
        pages = {L6},
          doi = {10.3847/2041-8213/abc550}}

@ARTICLE{2012MNRAS.419.2448P,
       author = {{Parker}, Richard J. and {Quanz}, Sascha P.},
        title = "{The effects of dynamical interactions on planets in young substructured star clusters}",
      journal = {\mnras},
         year = 2012,
        month = jan,
       volume = {419},
       number = {3},
        pages = {2448-2458},
          doi = {10.1111/j.1365-2966.2011.19911.x}}

@BOOK{2003gmbp.book.....H,
       author = {{Heggie}, Douglas and {Hut}, Piet},
        title = "{The Gravitational Million-Body Problem: A Multidisciplinary Approach to Star Cluster Dynamics}",
         year = 2003,
        publisher = {Cambridge University Press},
       adsurl = {https://ui.adsabs.harvard.edu/abs/2003gmbp.book.....H}
}

@ARTICLE{2023MNRAS.526.1987F,
       author = {{Flammini Dotti}, Francesco and {Capuzzo-Dolcetta}, R. and {Kouwenhoven}, M.~B.~N.},
        title = "{The dynamical evolution of protoplanetary discs and planets in dense star clusters}",
      journal = {\mnras},
         year = 2023,
        month = dec,
       volume = {526},
       number = {2},
        pages = {1987-1996},
          doi = {10.1093/mnras/stad2819}}

@ARTICLE{2011MNRAS.411..859M,
       author = {{Malmberg}, Daniel and {Davies}, Melvyn B. and {Heggie}, Douglas C.},
        title = "{The effects of fly-bys on planetary systems}",
      journal = {\mnras},
         year = 2011,
        month = feb,
       volume = {411},
       number = {2},
        pages = {859-877},
          doi = {10.1111/j.1365-2966.2010.17730.x}}

@ARTICLE{2013ApJ...766...81F,
       author = {{Fressin}, Fran{\c{c}}ois and {Torres}, Guillermo and {Charbonneau}, David and {Bryson}, Stephen T. and {Christiansen}, Jessie and {Dressing}, Courtney D. and {Jenkins}, Jon M. and {Walkowicz}, Lucianne M. and {Batalha}, Natalie M.},
        title = "{The False Positive Rate of Kepler and the Occurrence of Planets}",
      journal = {\apj},
         year = 2013,
        month = apr,
       volume = {766},
       number = {2},
          eid = {81},
        pages = {81},
          doi = {10.1088/0004-637X/766/2/81}}

@ARTICLE{2013MNRAS.433..867H,
       author = {{Hao}, W. and {Kouwenhoven}, M.~B.~N. and {Spurzem}, R.},
        title = "{The dynamical evolution of multiplanet systems in open clusters}",
      journal = {\mnras},
         year = 2013,
        month = jul,
       volume = {433},
       number = {1},
        pages = {867-877},
          doi = {10.1093/mnras/stt771}}

@ARTICLE{2018MNRAS.474.4460R,
       author = {{Ragusa}, Enrico and {Rosotti}, Giovanni and {Teyssandier}, Jean and {Booth}, Richard and {Clarke}, Cathie J. and {Lodato}, Giuseppe},
        title = "{Eccentricity evolution during planet-disc interaction}",
      journal = {\mnras},
         year = 2018,
        month = mar,
       volume = {474},
       number = {4},
        pages = {4460-4476},
          doi = {10.1093/mnras/stx3094}}

@ARTICLE{2020MNRAS.491.1369A,
       author = {{Anderson}, Kassandra R. and {Lai}, Dong and {Pu}, Bonan},
        title = "{In situ scattering of warm Jupiters and implications for dynamical histories}",
      journal = {\mnras},
         year = 2020,
        month = jan,
       volume = {491},
       number = {1},
        pages = {1369-1383},
          doi = {10.1093/mnras/stz3119}}

@ARTICLE{2023MNRAS.523.2832R,
       author = {{Romanova}, M.~M. and {Koldoba}, A.~V. and {Ustyugova}, G.~V. and {Lai}, D. and {Lovelace}, R.~V.~E.},
        title = "{Eccentricity growth of massive planets inside cavities of protoplanetary discs}",
      journal = {\mnras},
         year = 2023,
        month = aug,
       volume = {523},
       number = {2},
        pages = {2832-2849},
          doi = {10.1093/mnras/stad987}}

@ARTICLE{1984AJ.....89.1559H,
       author = {{Hills}, J.~G.},
        title = "{Close encounters between a star-planet system and a stellar intruder}",
      journal = {\aj},
         year = 1984,
        month = oct,
       volume = {89},
        pages = {1559-1564},
          doi = {10.1086/113659}}

@ARTICLE{2006ApJ...641..504A,
       author = {{Adams}, Fred C. and {Proszkow}, Eva M. and {Fatuzzo}, Marco and {Myers}, Philip C.},
        title = "{Early Evolution of Stellar Groups and Clusters: Environmental Effects on Forming Planetary Systems}",
      journal = {\apj},
         year = 2006,
        month = apr,
       volume = {641},
       number = {1},
        pages = {504-525},
          doi = {10.1086/500393}}

@ARTICLE{2017MNRAS.470.4337C,
       author = {{Cai}, Maxwell Xu and {Kouwenhoven}, M.~B.~N. and {Portegies Zwart}, Simon F. and {Spurzem}, Rainer},
        title = "{Stability of multiplanetary systems in star clusters}",
      journal = {\mnras},
         year = 2017,
        month = oct,
       volume = {470},
       number = {4},
        pages = {4337-4353},
          doi = {10.1093/mnras/stx1464}}

@ARTICLE{2019MNRAS.489.4311C,
       author = {{Cai}, Maxwell X. and {Portegies Zwart}, S. and {Kouwenhoven}, M.~B.~N. and {Spurzem}, Rainer},
        title = "{On the survivability of planets in young massive clusters and its implication of planet orbital architectures in globular clusters}",
      journal = {\mnras},
         year = 2019,
        month = nov,
       volume = {489},
       number = {3},
        pages = {4311-4321},
          doi = {10.1093/mnras/stz2467}}

@ARTICLE{2020MNRAS.497.1807S,
       author = {{Stock}, Katja and {Cai}, Maxwell X. and {Spurzem}, Rainer and {Kouwenhoven}, M.~B.~N. and {Portegies Zwart}, Simon},
        title = "{On the survival of resonant and non-resonant planetary systems in star clusters}",
      journal = {\mnras},
         year = 2020,
        month = sep,
       volume = {497},
       number = {2},
        pages = {1807-1825},
          doi = {10.1093/mnras/staa2047}}

@ARTICLE{2015MNRAS.449.3543W,
       author = {{Wang}, Long and {Kouwenhoven}, M.~B.~N. and {Zheng}, Xiaochen and {Church}, Ross P. and {Davies}, Melvyn B.},
        title = "{Close encounters involving free-floating planets in star clusters}",
      journal = {\mnras},
         year = 2015,
        month = jun,
       volume = {449},
       number = {4},
        pages = {3543-3558},
          doi = {10.1093/mnras/stv542}}

@ARTICLE{2022MNRAS.509.5253W,
       author = {{Wang}, Yi-Han and {Perna}, Rosalba and {Leigh}, Nathan W.~C. and {Shara}, Michael M.},
        title = "{Hot Jupiter formation in dense clusters: secular chaos in multiplanetary systems}",
      journal = {\mnras},
         year = 2022,
        month = feb,
       volume = {509},
       number = {4},
        pages = {5253-5264},
          doi = {10.1093/mnras/stab3321}}

@ARTICLE{2022MNRAS.512.2460S,
       author = {{Stock}, Katja and {Veras}, Dimitri and {Cai}, Maxwell X. and {Spurzem}, Rainer and {Portegies Zwart}, Simon},
        title = "{Birth cluster simulations of planetary systems with multiple super-Earths: initial conditions for white dwarf pollution drivers}",
      journal = {\mnras},
         year = 2022,
        month = may,
       volume = {512},
       number = {2},
        pages = {2460-2473},
          doi = {10.1093/mnras/stac602}}

@ARTICLE{2015ApJ...799..180S,
       author = {{Silburt}, Ari and {Gaidos}, Eric and {Wu}, Yanqin},
        title = "{A Statistical Reconstruction of the Planet Population around Kepler Solar-type Stars}",
      journal = {\apj},
         year = 2015,
        month = feb,
       volume = {799},
       number = {2},
          eid = {180},
        pages = {180},
          doi = {10.1088/0004-637X/799/2/180}}

@ARTICLE{2024MNRAS.533.3484S,
       author = {{Schoettler}, Christina and {Owen}, James E.},
        title = "{The effect of dynamical interactions in stellar birth environments on the orbits of young close-in planetary systems}",
      journal = {\mnras},
         year = 2024,
        month = sep,
       volume = {533},
       number = {3},
        pages = {3484-3500},
          doi = {10.1093/mnras/stae1900}
}

@ARTICLE{2025arXiv250106342V,
       author = {{Van Zandt}, Judah and {Petigura}, Erik A. and {Lubin}, Jack and {Weiss}, Lauren M. and {Turtelboom}, Emma V. and {Fetherolf}, Tara and {Akana Murphy}, Joseph M. and {Crossfield}, Ian J.~M. and {Gilbert}, Greg and {Mocnik}, Teo and {Batalha}, Natalie M. and {Dressing}, Courtney and {Fulton}, Benjamin and {Howard}, Andrew W. and {Huber}, Daniel and {Isaacson}, Howard and {Kane}, Stephen R. and {Robertson}, Paul and {Roy}, Arpita and {Angelo}, Isabel and {Behmard}, Aida and {Beard}, Corey and {Chontos}, Ashley and {Dai}, Fei and {Dalba}, Paul A. and {Giacalone}, Steven and {Hill}, Michelle L. and {Hirsch}, Lea A. and {Holcomb}, Rae and {Howell}, Steve B. and {Mayo}, Andrew W. and {MacDougall}, Mason G. and {Pidhorodetska}, Daria and {Polanski}, Alex S. and {Rogers}, James and {Rosenthal}, Lee J. and {Rubenzahl}, Ryan A. and {Scarsdale}, Nicholas and {Tyler}, Dakotah and {Yee}, Samuel W. and {Zink}, Jon},
        title = "{The TESS-Keck Survey XXIV: Outer Giants may be More Prevalent in the Presence of Inner Small Planets}",
      journal = {arXiv e-prints},
         year = 2025,
        month = jan,
          eid = {arXiv:2501.06342},
        pages = {arXiv:2501.06342},
archivePrefix = {arXiv},
       eprint = {2501.06342}}



\appendix

\section{Additional plots}

\begin{figure*}
    \centering
    \begin{minipage}[t]{1.0\columnwidth}
         \centering
    	\includegraphics[width=1.0\linewidth]{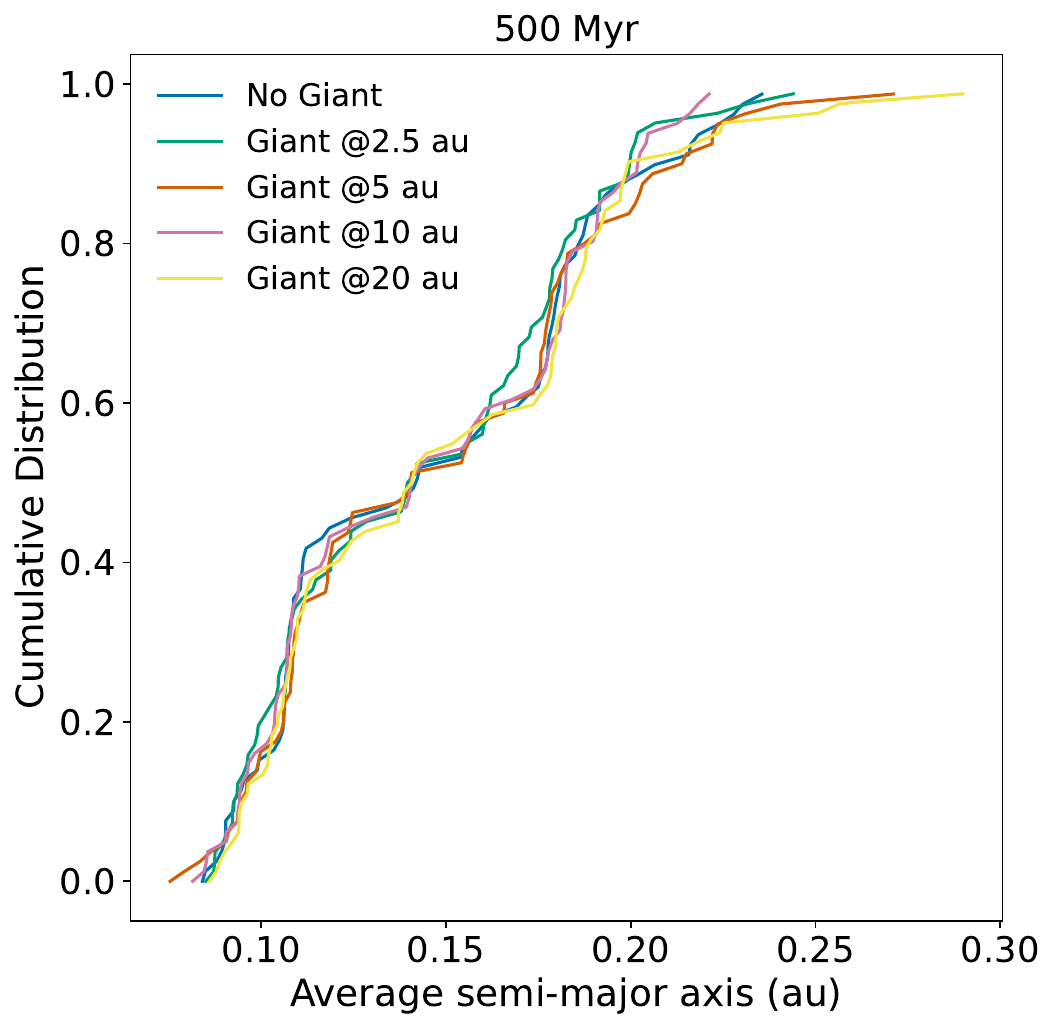}
     \end{minipage}
        \centering
        \vspace{0pt}
    \begin{minipage}[t]{0.97\columnwidth}
    	\includegraphics[width=1.0\linewidth]{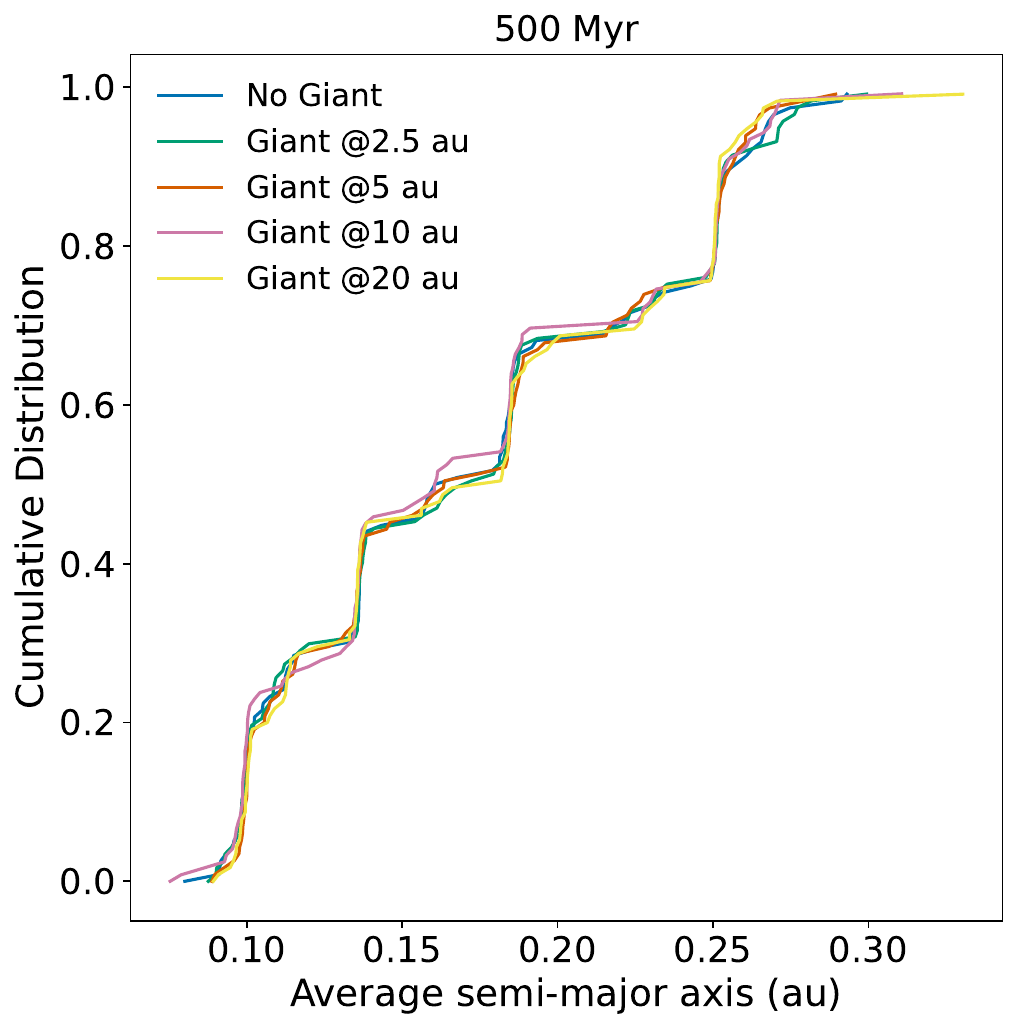}
      \end{minipage}
     \caption{Differences in semi-major axis of remaining close-in planets at 500 Myr for the higher-mass giant systems. These two plots (left: 10 R$_{\rm{MH}}$ separation, right: 14 R$_{\rm{MH}}$ separation) show the cumulative distributions of the semi-major axes of all remaining close-in planets at 500 Myr. There are only minor differences in the distributions, indicating that for each of the four placements of the distant giant planets (as well as no giant), the close-in planets show similar locations after they all experienced a fly-by. There are no obvious differences between the architectures, and even the absence of the giant planet produces a similar distribution in the semi-major axes at 500 Myr.}
     \label{fig:CDF_all_semax}
\end{figure*}

\begin{figure*}
    \centering
    \begin{minipage}[t]{1.0\columnwidth}
         \centering
    	\includegraphics[width=1.0\linewidth]{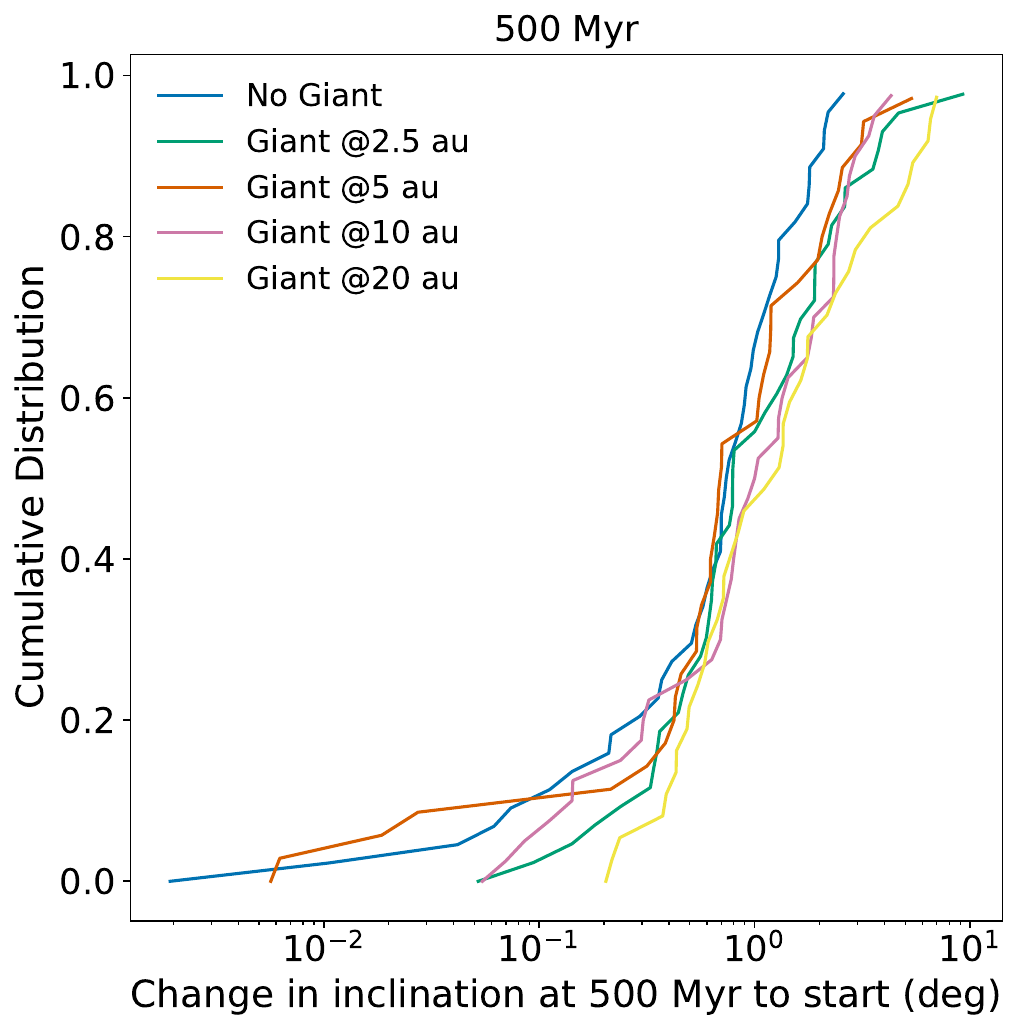}
     \end{minipage}
        \centering
        \vspace{0pt}
    \begin{minipage}[t]{1.0\columnwidth}
    	\includegraphics[width=1.0\linewidth]{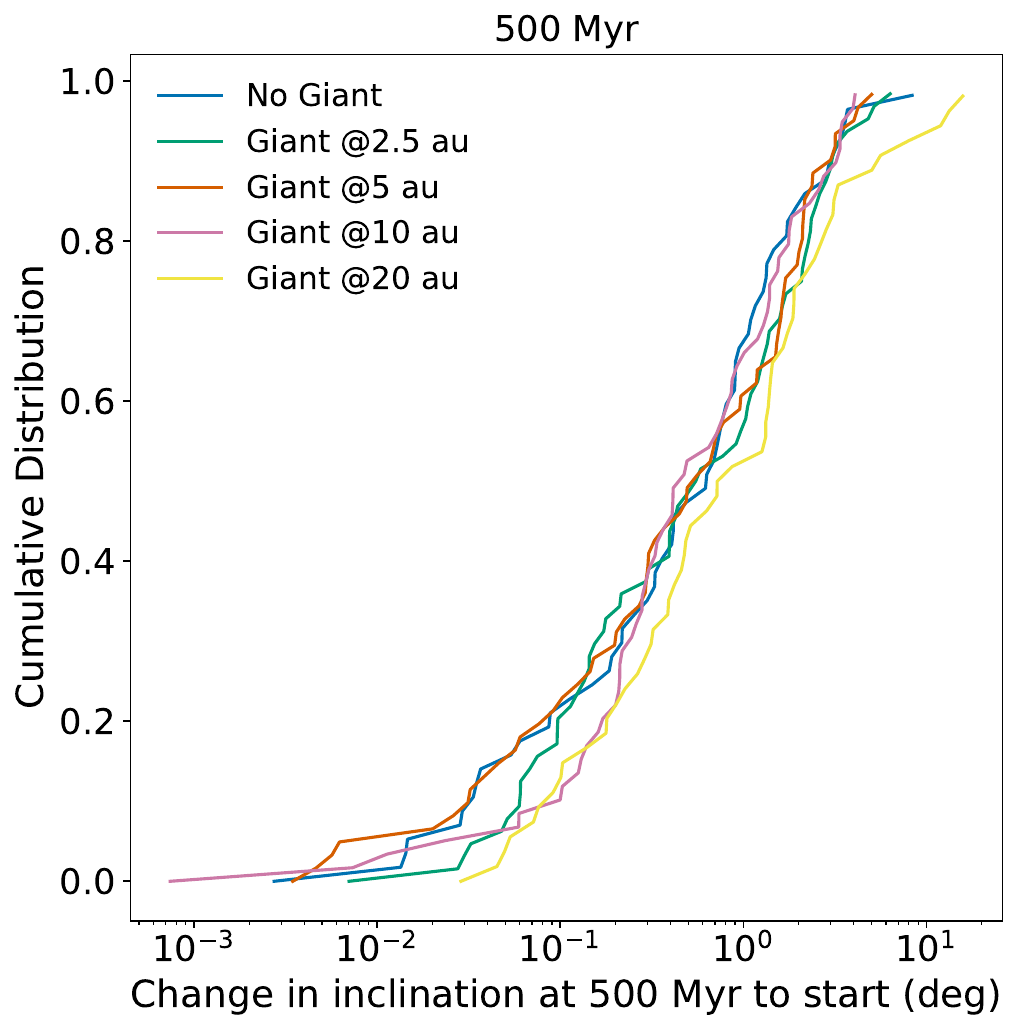}
      \end{minipage}
     \caption{Differences in inclination of remaining close-in planets at 500 Myr for the higher-mass giant systems. These two plots (left: 10 R$_{\rm{MH}}$ separation, right: 14 R$_{\rm{MH}}$ separation) show the cumulative distributions of the inclination of the remaining close-in planets at 500 Myr compared to the start of the simulations before the fly-by. There are only minor differences in the distributions, indicating that for each of the four placements of the distant giant planets (as well as no giant), the close-in planets show similar differences in their inclination after they all experienced a fly-by. There are no obvious differences between the architectures, and even the absence of the giant planet produces a similar distribution at 500 Myr.}
     \label{fig:CDF_all_incl}
\end{figure*}


\begin{figure*}
    \centering
         \centering
    	\includegraphics[width=0.9\linewidth]{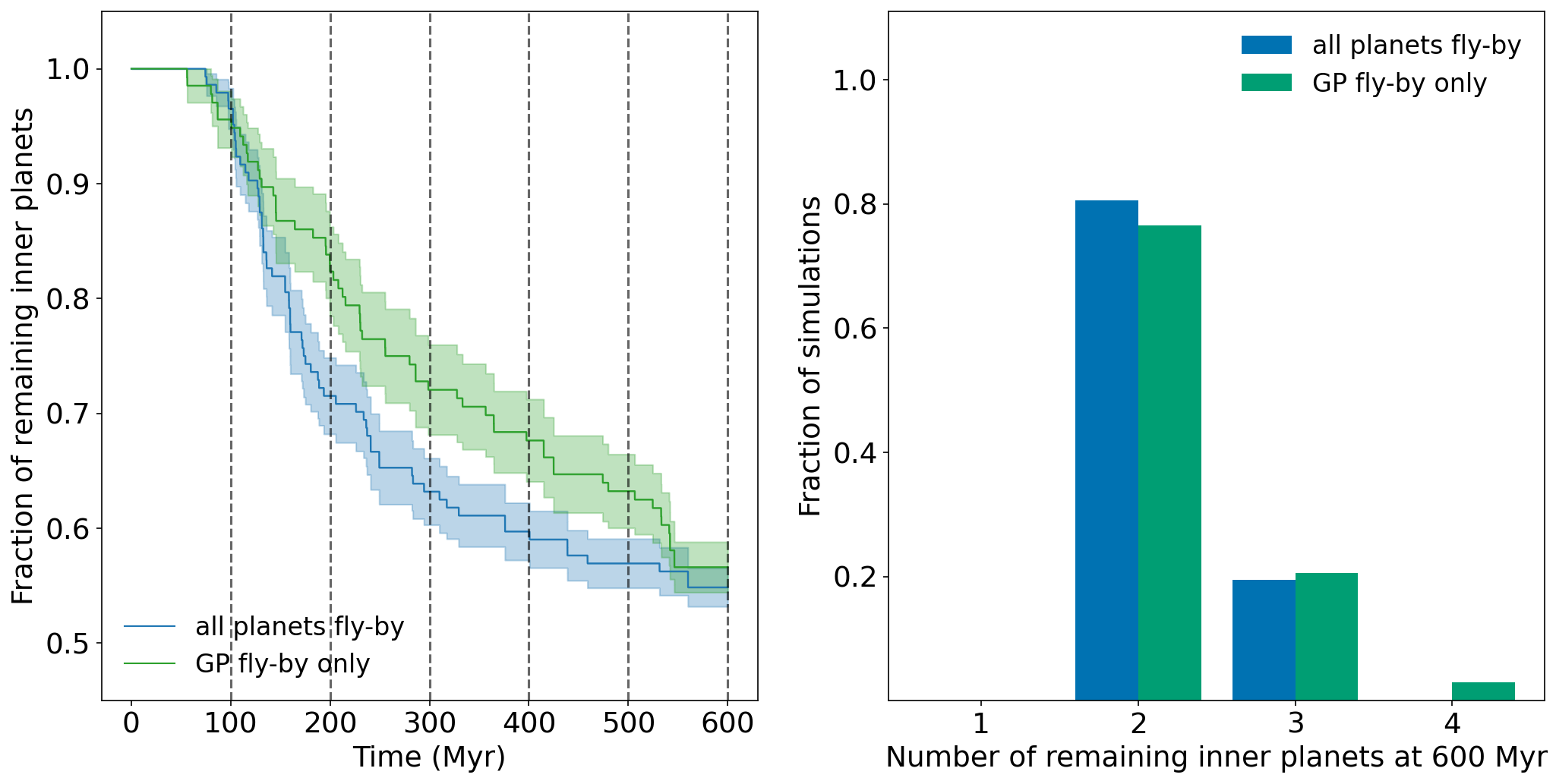}
        \centering
        \vspace{0pt}
    	\includegraphics[width=0.9\linewidth]{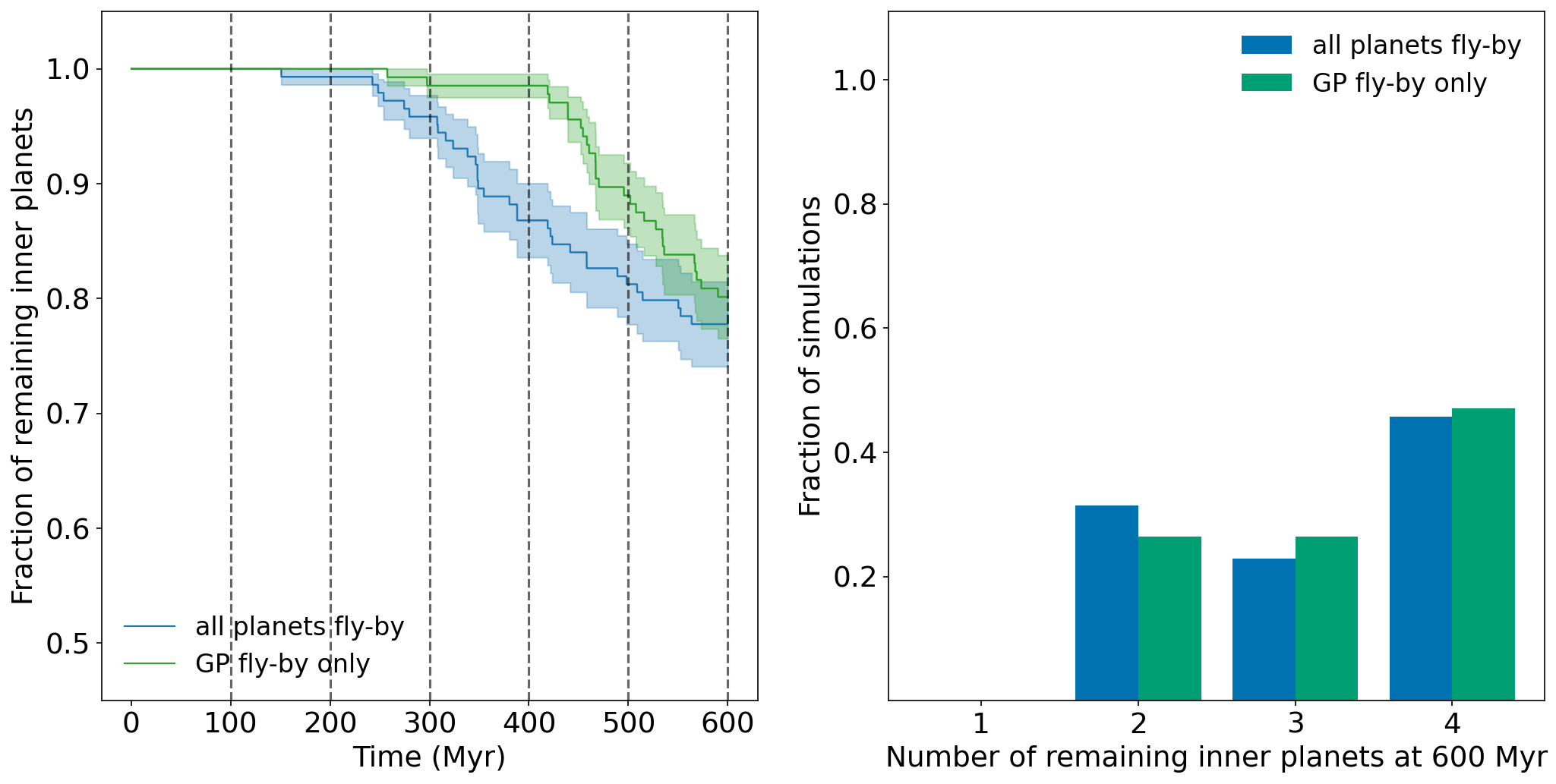}
     \caption{Differences in remaining number/fraction close-in planets at an extended time of 600 Myr for the 20 au giant locations. The extended evolution of the number of remaining planets in the systems with a 20 au giant planet location is shown (top: 10 R$_{\rm{MH}}$ separation, bottom: 14 R$_{\rm{MH}}$ separation). This has been run for an additional 100 Myr to 600 Myr and shows that the difference in the average number of inner planets between the externally and internally perturbed cases for the 20 au giant planet distance converges over 600 Myr with the same end state.}
     \label{fig:CDF_20_au_ext}
\end{figure*}


\begin{figure*}
    \centering
         \centering
    	\includegraphics[width=0.6\linewidth]{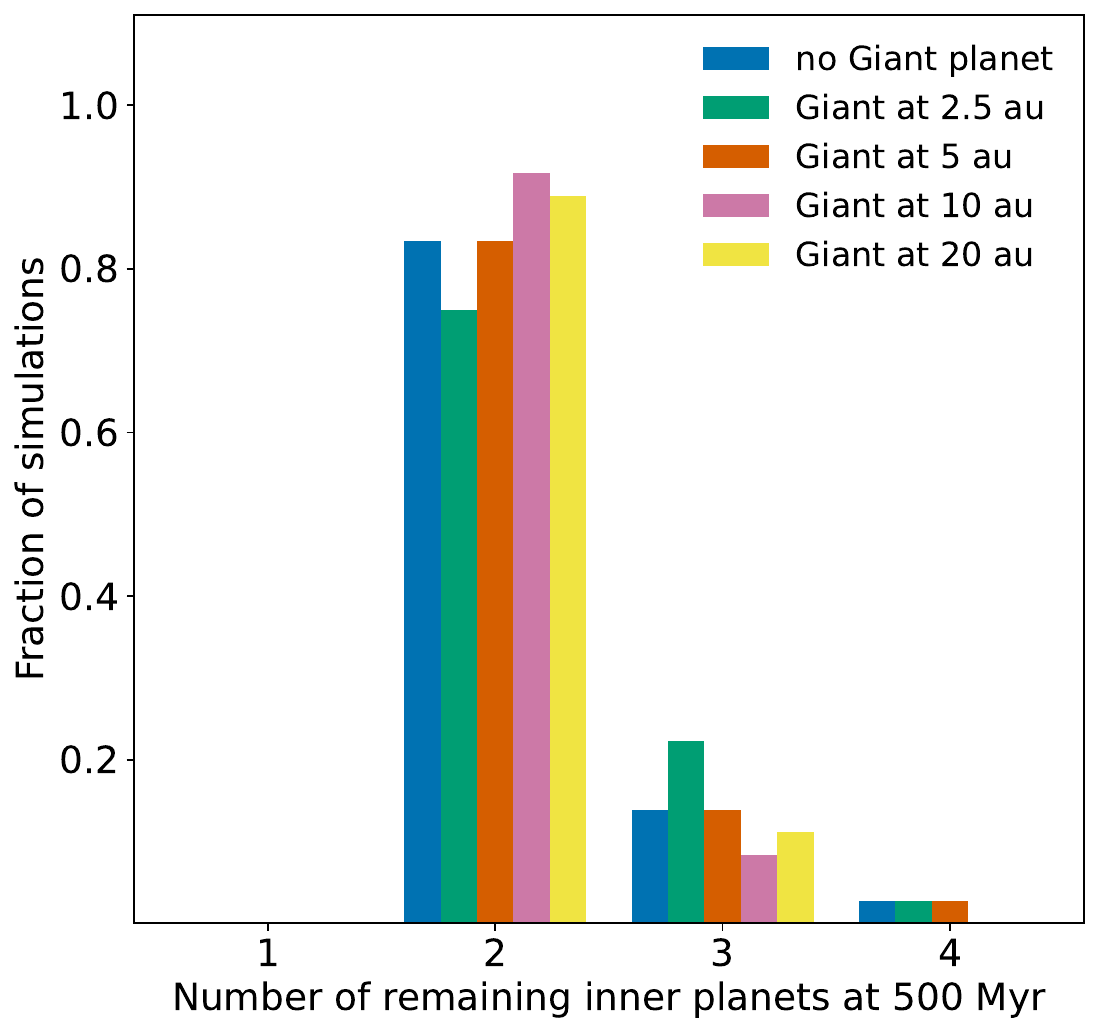}
        \centering
        \vspace{0pt}
    	\includegraphics[width=0.6\linewidth]{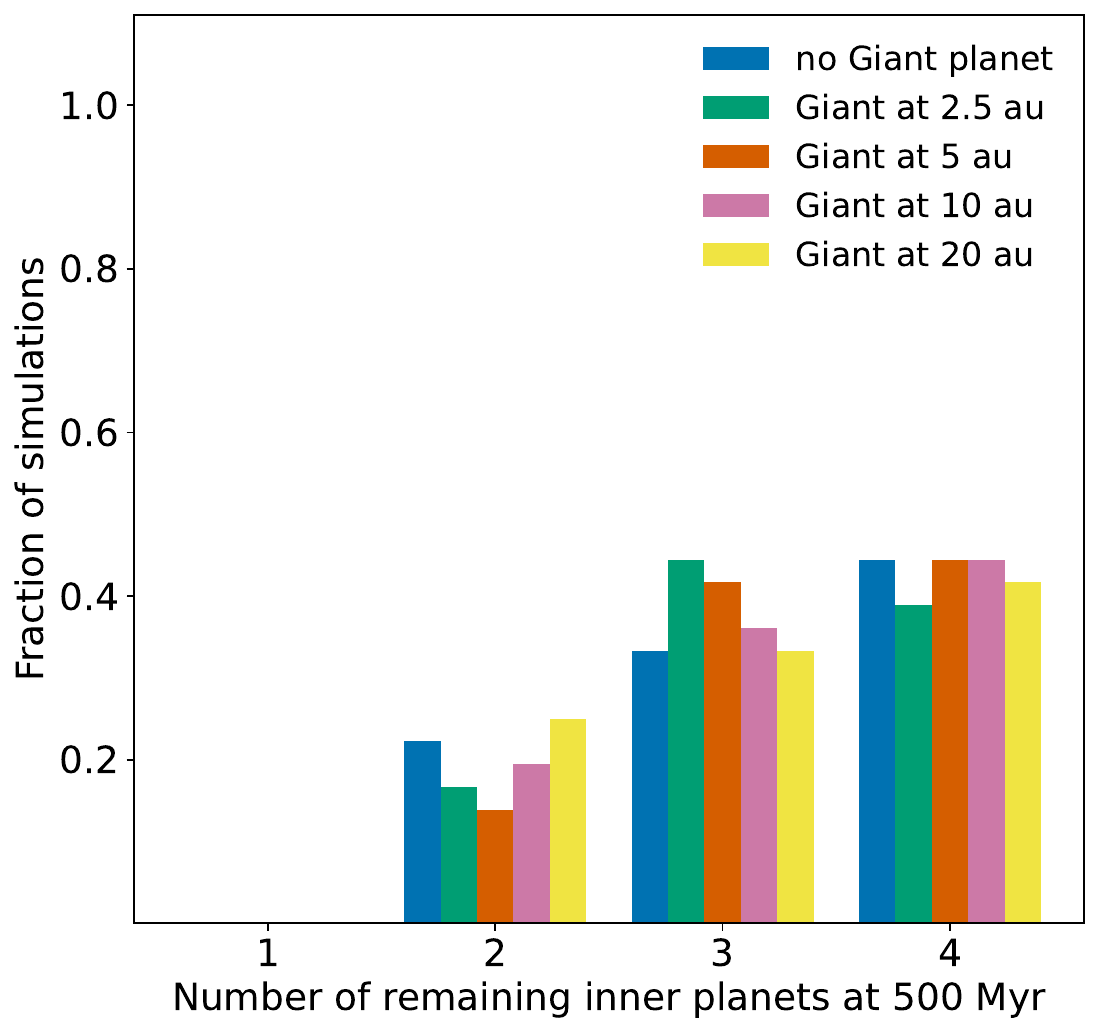}
     \caption{Number of planets for simulations with a lower-mass giant planet at 500 Myr. We show the histogram with the comparison between the externally perturbed inner-planet systems for the initial 10 R$_{\rm{MH}}$ (top) and 14 R$_{\rm{MH}}$ (bottom) separation systems for all giant planet distances with a lower giant planet mass. Also plotted is the number of inner planets in systems where the Giant planet is completely absent. \textbf{Top:} We find that, as in the higher giant mass case, there is no significant difference between the average planet numbers, regardless of distance and the absence/presence of the giant. At 500 Myr, most inner planet systems have moved from a four-planet to a two-planet setup, similar to the higher mass giant scenario.\textbf{Bottom:} As in the higher giant mass case, there is no significant difference between the average planet numbers, regardless of distance and the absence/presence of the giant. At 500 Myr, the inner planet systems show a mix of inner planet numbers, split evenly between four- and three-planet systems, with a smaller percentage of two-planet systems.}
     \label{fig:low_mass_hist}
\end{figure*}

\begin{figure*}
    \centering
         \centering
    	\includegraphics[width=0.95\linewidth]{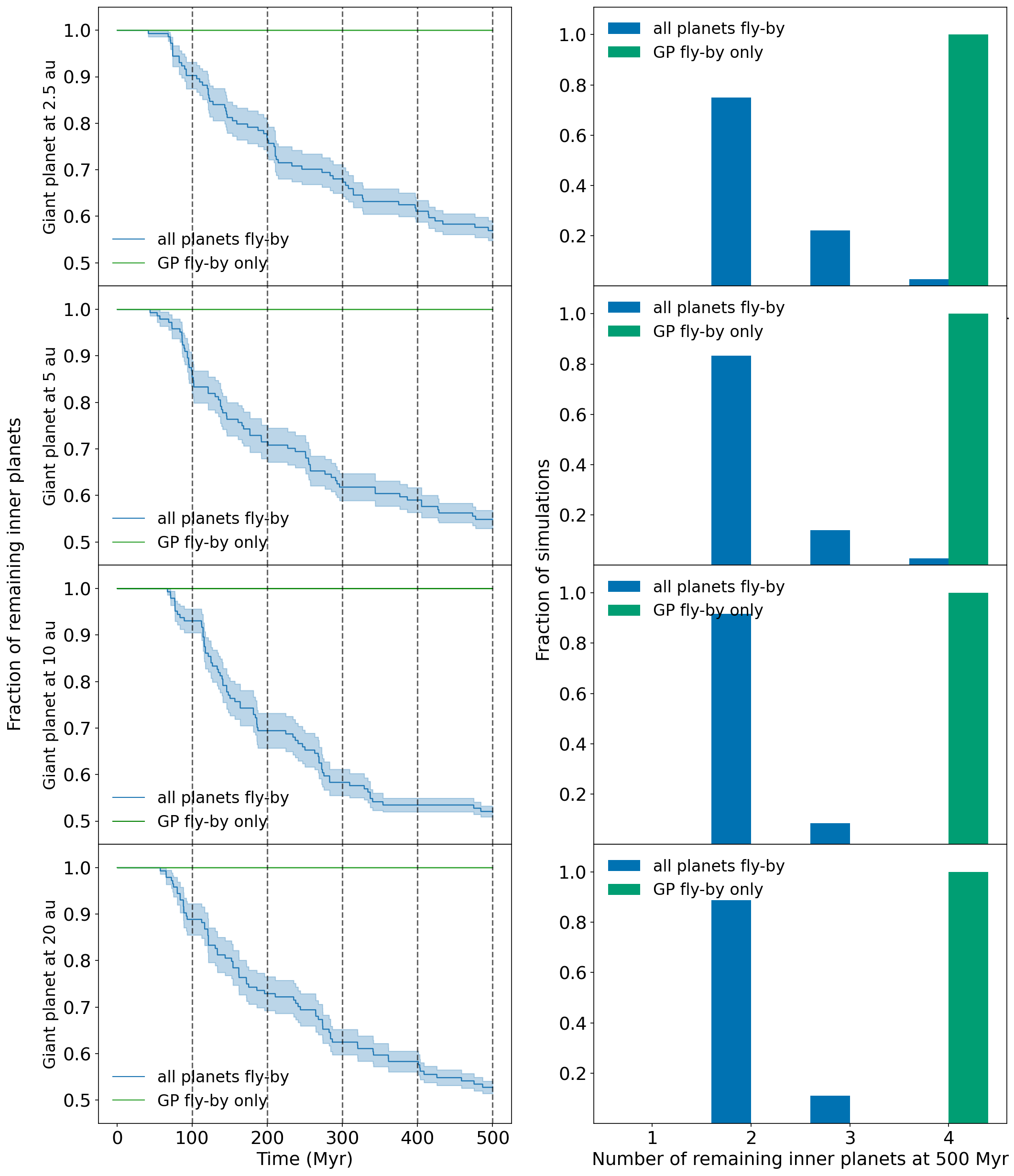}
        \centering
        \vspace{0pt}
    \centering
      \caption{In this graph, we show the difference in the evolution of the inner planet numbers for the 10 R$_{\rm{MH}}$ simulations for the lower giant planet mass, comparing the external and internal perturbation case. We find that while the externally perturbed systems show collisions after $\sim$50 Myr, the internally perturbed systems show no collisions at all over the 500 Myr simulation time.}
     \label{fig:plan_num_lowmass_10_int}
\end{figure*}

\begin{figure*}
         \centering
    	\includegraphics[width=0.95\linewidth]{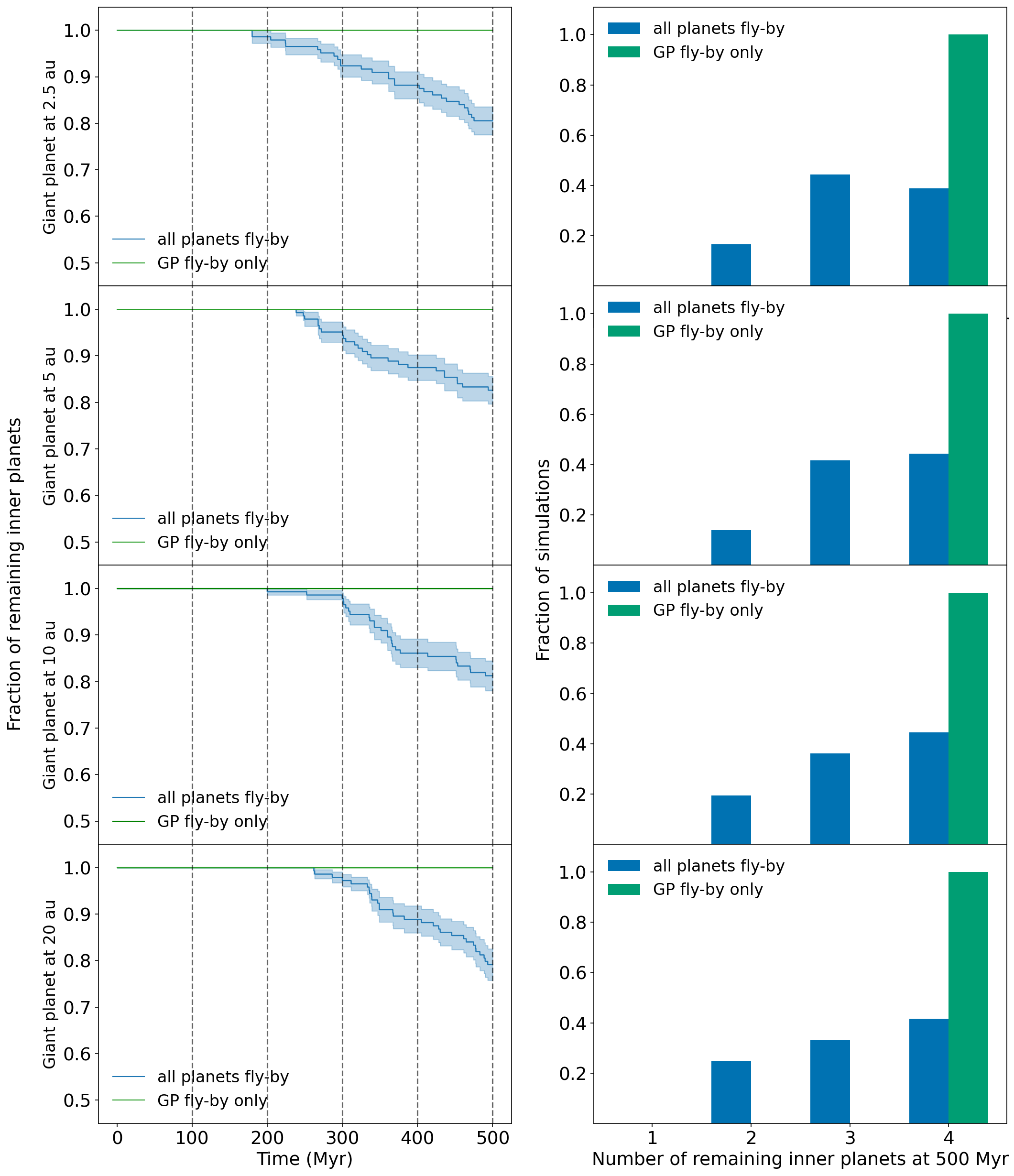}
        \centering
        \vspace{0pt}
     \caption{In this graph, we show the difference in the evolution of the inner planet numbers for the 14 R$_{\rm{MH}}$ simulations for the lower giant planet mass, comparing the external and internal perturbation case. We find that while the externally perturbed systems show collisions after $\sim$200--250 Myr, the internally perturbed systems show no collisions at all over the 500 Myr simulation time.}
     \label{fig:plan_num_lowmass_14_int}
\end{figure*}

\bsp	
\label{lastpage}
\end{document}